# Design of Learning Based MIMO Cognitive Radio Systems

Feifei Gao, Rui Zhang, Ying-Chang Liang, and Xiaodong Wang

**Abstract**

This paper addresses the design issues of the multi-antenna-based cognitive radio (CR) system that is able to operate concurrently with the licensed primary radio (PR) system. We propose a practical CR transmission strategy consisting of three major stages: environment learning, channel training, and data transmission. In the environment learning stage, the CR transceivers both listen to the PR transmission and apply blind algorithms to estimate the spaces that are orthogonal to the channels from the PR. Assuming time-division duplex (TDD) based transmission for the PR, cognitive beamforming is then designed and applied at CR transceivers to restrict the interference to/from the PR during the subsequent channel training and data transmission stages. In the channel training stage, the CR transmitter sends training signals to the CR receiver, which applies the linear-minimum-mean-square-error (LMMSE) based estimator to estimate the effective channel. Considering imperfect estimations in both learning and training stages, we derive a lower bound on the ergodic capacity achievable for the CR in the data transmission stage. From this capacity lower bound, we observe a general learning/training/throughput tradeoff associated with the proposed scheme, pertinent to transmit power allocation between training and transmission stages, as well as time allocation among learning, training, and transmission stages. We characterize the aforementioned tradeoff by optimizing the associated power and time allocation to maximize the CR ergodic capacity.

**Index Terms**

Cognitive radio, spectrum sharing, multi-antenna systems, environment learning, channel training.

F. Gao is with the School of Engineering and Science, Jacobs University Bremen, Campus Ring 1, Bremen, Germany, 28759 (Email: feifeigao@ieee.org).

R. Zhang and Y.-C. Liang are with the Institute for Infocomm Research, A*STAR, 1 Fusionopolis Way, #21-01 Connexis, Singapore 138632 (Email: {rzhang, ycliang}@i2r.a-star.edu.sg).

X. Wang is with the Department of Electrical Engineering, Columbia University, New York, USA, (Email: wangx@ee.columbia.edu).

# I. Introduction

The original idea of cognitive radio (CR) envisions that the CR opportunistically accesses the frequency bands allocated to the licensed primary radio (PR) system when the latter is not in operation [1]. In particular, the CR first detects the void frequency bands, also known as "frequency holes", and then transmits over them. The related key technique is called *spectrum sensing*, which has been thoroughly studied in the literatures over the recent years [2]–[5]. This *opportunistic spectrum access* (OSA) idea for the CR has been proven meaningful from the survey made by the Federal Communications Commission (FCC) [6], which reveals that the current utilization efficiency of the licensed radio spectrums could be as low as 15% on average. An alternative model for the operation of the CR other than OSA is known as *spectrum sharing* (SS) [7], for which the concurrent transmission of CR and PR in the same frequency band is permissible provided that the resultant interference power due the the CR transmission at each PR terminal, or the so-called *interference temperature* (IT), is kept below a predefined threshold.

A new type of SS transmission scheme was recently proposed in [8], where multiple antennas are deployed at the CR transmitter (CR-Tx) to enable *cognitive beamforming* for regulating the resultant interference power levels at PR terminals. However, the scheme proposed in [8] requires perfect knowledge of all the channels from CR-Tx to PR terminals available at CR-Tx. This assumption is not realistic from a practical viewpoint since the PR is in general not responsible to facilitate the CR in obtaining such channel knowledge. Under the assumption of time-division duplex (TDD) transmission mode for the PR, a breakthrough was made later in [9], where a blind estimation approach is proposed for CR-Tx to obtain partial channel information from CR-Tx to PR terminals. Based on the estimated partial channel information, transmit cognitive beamforming is designed and is shown to be capable of directing CR's transmit signals only through the null space of the CR-PR channels and thus removing the interference to PR terminals. Unfortunately, this very initial effort made in [9] is still far from pushing this SS scheme into practical usage; for example, the channels between CR transceivers are assumed perfect and the interference from PR to CR terminals is ignored for the CR transmission design.

In this work, we develop a more practical CR transmission strategy, where many issues that were not addressed in [9] are embraced. The main contributions are summarized as follows:

- The proposed CR transmission scheme consists of three major stages: environmental learning, channel training, and data transmission. Note that this new scheme is more concrete as well as practical in comparison with that in the existing work [9].
- In addition to the transmit cognitive beamforming method studied in [9], we propose a new beamforming method at the CR receiver (CR-Rx) to mitigate the interference from the PR. More specifically, both CR-Tx and CR-Rx listen to the PR transmission during the environment learning stage and then design the transmit and receive beamforming to null the interference to and from the PR, respectively.
- Instead of assuming perfect channel knowledge between CR-Tx and CR-Rx as in [9], we adopt a training stage for the CR to estimate the effective channel after applying joint transmit and receive beamforming. The optimal training structure is derived to minimize the channel estimation error, taking into account of the interferences to and from the PR.
- We derive a lower bound on the ergodic capacity achievable for the CR in the data transmission stage, subject to a prescribed IT constraint at the PR, from which we observe a new *learning/training/throughput tradeoff* [1] associated with the proposed CR transmission scheme, pertinent to transmit power allocation between training and data transmission stages, as well as time allocation among learning, training, and data transmission stages. Moreover, we optimize the associated power and time allocation to maximize the derived lower bound of the CR ergodic capacity.

The rest of this paper is organized as follows. Section II presents the system model of the multiple-antenna CR system. Section III formulates the CR learning, training, and transmission strategies. Section IV derives the lower bound on the CR ergodic capacity, and obtains the optimal power and time allocation among different stages to maximize this lower bound. Section V provides simulation results to corroborate the proposed studies. Finally, Section VI concludes the paper.

**Notations:** Vectors and matrices are boldface small and capital letters, respectively; the transpose, complex conjugate, Hermitian, inverse, and pseudo-inverse of a matrix $\mathbf{A}$ are denoted by

[1]This tradeoff is more general as well as of more practical relevance than the earlier proposed sensing-throughput tradeoff [10] and learning-throughput tradeoff [9] for OSA- and SS-based CR systems, respectively.

$\mathbf{A}^T$, $\mathbf{A}^*$, $\mathbf{A}^H$, $\mathbf{A}^{-1}$, and $\mathbf{A}^\dagger$, respectively; $tr(\mathbf{A})$ and $\det(\mathbf{A})$ denote the trace and the determinant of the matrix $\mathbf{A}$, respectively; $\mathrm{diag}\{\mathbf{a}\}$ is a diagonal matrix whose diagonal elements are given by entries of the vector $\mathbf{a}$; $\mathbf{I}$ denotes the identity matrix; and $\mathrm{E}[\cdot]$ denotes the statistical expectation.

## II. System Model

We consider a CR system with $M_1$ antennas at terminal CR-T1 and $M_2$ antennas at terminal CR-T2,[2] as shown in Fig. 1. At the same operating frequency band, there exists a PR link with two terminals PR-T1 and PR-T2. We assume a time-division-duplex (TDD) mode for both PR and CR links. Specifically, the transmitting of PR-T1 occupies an average proportion $\alpha$ of the overall period, while its receiving occupies the other $(1-\alpha)$ of the overall period. For simplicity, we assume that PR-T2 stays outside the CR's transmission boundary, as shown in Fig. 1. Nevertheless, all the following discussions can be straightforwardly extended to considering both PR-T1 and PR-T2 inside the CR's boundary by utilizing the *effective interference channel* concept proposed in [9]. We then denote the number of antennas at PR-T1 as $M_p$ and replace PR-T1 by PR for notational brevity.

Let the channels from PR to CR-T1 and CR-T2 be represented by the $M_1 \times M_p$ matrix $\mathbf{G}_1$ and the $M_2 \times M_p$ matrix $\mathbf{G}_2$, respectively. The channel from CR-T1 to CR-T2 is denoted by the $M_2 \times M_1$ matrix $\mathbf{H}$. Each element of all the channels involved is assumed to be independent circularly symmetric complex Gaussian (CSCG) random variable with zero mean and unit variance. Since both PR and CR operate in a TDD mode, the channel reciprocity principle is justifiable and thus the reverse channels from CR-T1 to PR, from CR-T2 to PR, and from CR-T2 to CR-T1 are assumed to be $\mathbf{G}_1^T$, $\mathbf{G}_2^T$, and $\mathbf{H}^T$, respectively. Furthermore, we require more antennas at CRs than at PR, i.e., $M_j > M_p$ for $j = 1, 2$, in order to enable the environment learning method discussed later in this paper. This requirement on the number of CR's antennas is a reasonable cost for the CR to realize the concurrent transmission with PR.

## III. CR Transmission Strategy

As shown in Fig. 2, the CR transmission is divided into consecutive frames, each having a duration of $N$ symbol periods. Each frame is further divided into three consecutive stages:

[2]We do not specify CR-Tx or CR-Rx because both CR terminals transmit and receive alternately in a TDD mode.

environment learning, channel training, and data transmission with durations of $N_l$, $N_t$, and $N_d$ symbol periods, respectively. Obviously, there is $N_l + N_t + N_d = N$. In the environment learning stage, CR-T1 and CR-T2 gain partial knowledge on $\mathbf{G}_1$ and $\mathbf{G}_2$ via listening to the PR's transmission. Since this knowledge is obtained in a passive manner, we describe it with the term "learning". In contrast, in the second channel training stage, the CR transmitter actively sends out training signals for the receiver to estimate the channel between CR-T1 and CR-T2, and thus, this process is described by the term "training". During the last data transmission stage, CR-T1 and CR-T2 transmit in an alternate manner. Note that the value of $N$ is chosen to be, on one hand, sufficiently smaller than the channel coherence time such that all the channels can be safely assumed to be constant within each frame, and on the other hand, as large as possible in order to save the overall throughput loss due to learning and training overheads.

### *A. Environment Learning Stage*

Considering that PR switches between transmitting and receiving, signals sent from PR can be expressed as

$$\mathbf{s}_p(n) = \begin{cases} \tilde{\mathbf{s}}_p(n) & \text{if PR transmits} \\ \mathbf{0} & \text{otherwise} \end{cases} \qquad n = 1 \ldots, N, \tag{1}$$

where $\tilde{\mathbf{s}}_p(n)$'s are independent and identically distributed (i.i.d.) random signals with covariance matrix $\sigma_s^2\mathbf{I}$. Then, the average covariance matrix over the entire time period is $\mathbf{R}_p = \mathrm{E}[\mathbf{s}_p(n)\mathbf{s}_p^H(n)] = \alpha\sigma_s^2\mathbf{I}$.

The signals received at CR-T1 and CR-T2 during the learning stage are then

$$\mathbf{y}_j(n) = \mathbf{G}_j\mathbf{s}_p(n) + \mathbf{z}_j(n), \quad n = 1, \ldots, N_l, \tag{2}$$

for $j = 1, 2$, where $\mathbf{z}_j(n)$ is the independent CSCG noise vector with zero means and the covariance matrix $\sigma_{nj}^2\mathbf{I}$.

*1) Ideal Case:* The covariance matrices of the received signals at CRs can be expressed as

$$\mathbf{R}_j = \mathrm{E}[\mathbf{y}_j(n)\mathbf{y}_j^H(n)] = \underbrace{\alpha\sigma_s^2\mathbf{G}_j\mathbf{G}_j^H}_{\mathbf{Q}_j} + \sigma_{nj}^2\mathbf{I}, \tag{3}$$

where $\mathbf{Q}_j$ is defined correspondingly. The eigen-value decomposition (EVD) of $\mathbf{R}_j$ is

$$\mathbf{R}_j = \mathbf{V}_j\mathbf{\Sigma}_j\mathbf{V}_j^H + \sigma_{nj}^2\mathbf{U}_j\mathbf{U}_j^H, \qquad j = 1, 2, \tag{4}$$

where $\mathbf{V}_j$ is the $M_j \times M_p$ signal subspace matrix and $\mathbf{U}_j$ is the $M_j \times (M_j - M_p)$ noise subspace matrix. Correspondingly, $\mathbf{\Sigma}_j$ is the diagonal matrix that contains the largest $M_p$ eigenvalues of $\mathbf{R}_j$. It is easy to verify that $\mathbf{U}_j^H \mathbf{G}_j = \mathbf{0}$ and $\mathbf{V}_j(\mathbf{\Sigma}_j - \sigma_{nj}^2 \mathbf{I})\mathbf{V}_j^H = \mathbf{Q}_j$.

The channel $\mathbf{G}_j$ is related to $\mathbf{V}_j$ by $\mathbf{G}_j = \mathbf{V}_j \mathbf{B}_j$, where $\mathbf{B}_j$ is an unknown $M_p \times M_p$ matrix. Fortunately, knowing $\mathbf{V}_j$ and $\mathbf{U}_j$ is sufficient to design the *cognitive transmit beamforming* [9]. That is, CR terminals transmit only through the space spanned by $\mathbf{U}_j^*$, thereby no interference is caused to PR because $\mathbf{G}_j^T \mathbf{U}_j^* = \mathbf{0}$. Therefore, the main task for CR-T1 (CR-T2) in the learning stage is to blindly estimate the noise subspace matrix $\mathbf{U}_1$ ($\mathbf{U}_2$) from the received signal covariance matrix, $\mathbf{R}_1$ ($\mathbf{R}_2$).

*2) Practical Case:* Given the finite number of samples received from PR, the sample covariance matrix for the received signals at each CR terminal is computed as

$$\hat{\mathbf{R}}_j = \frac{1}{N_l} \sum_{n=1}^{N_l} \mathbf{y}_j(n) \mathbf{y}_j^H(n), \quad j = 1, 2. \tag{5}$$

The EVD of $\hat{\mathbf{R}}_j$ is written as

$$\hat{\mathbf{R}}_j = \hat{\mathbf{V}}_j \hat{\mathbf{\Sigma}}_j \hat{\mathbf{V}}_j^H + \hat{\mathbf{U}}_j \hat{\mathbf{\Gamma}} \hat{\mathbf{U}}_j^H. \tag{6}$$

From [11], the first-order perturbation of the noise subspace due to the finite received samples can be approximated by

$$\Delta \mathbf{U}_j = \hat{\mathbf{U}}_j - \mathbf{U}_j \approx -\mathbf{Q}_j^\dagger \Delta \mathbf{R}_j \mathbf{U}_j, \tag{7}$$

where $\Delta \mathbf{R}_j \triangleq \hat{\mathbf{R}}_j - \mathbf{R}_j$.

### *B. Data Transmission Stage*

Before we make discussions for the channel training stage, we need to first recognize the required channels for data detection at both CR terminals. Thus, we bring forward the discussions for the data transmission stage here.

Suppose that on average CR-T1 transmits over $\theta N_d$ symbol periods whose indices belong to the set $\mathcal{N}_{d1}$ and CR-T2 transmits over the remaining $(1-\theta)N_d$ symbol periods whose indices belong to the set $\mathcal{N}_{d2}$, where $\theta \le 1$ is a prescribed constant. Note that $\mathcal{N}_{d1} \bigcup \mathcal{N}_{d2} = \{N_l + N_t + 1, N_l + N_t + 2 \ldots, N-1, N\}$ and $\mathcal{N}_{d1} \bigcap \mathcal{N}_{d2} = \emptyset$. Denote the encoded signal vector from CR-T1 and CR-T2 at symbol period $n$ as $\mathbf{d}_1(n)$ and $\mathbf{d}_2(n)$, respectively. We look into the following two cases:

*1) Ideal Case:* To protect PR, $\mathbf{d}_j(n)$ is precoded by $\mathbf{U}_j^*$ from the earlier introduced cognitive transmit beamforming. The received signals at CR-T1 and CR-T2 are

$$\mathbf{y}_1(n) = \mathbf{H}^T\mathbf{U}_2^*\mathbf{d}_2(n) + \mathbf{G}_1\mathbf{s}_p(n) + \mathbf{z}_1(n), \qquad n \in \mathcal{N}_{d2}, \tag{8a}$$

$$\mathbf{y}_2(n) = \mathbf{H}\mathbf{U}_1^*\mathbf{d}_1(n) + \mathbf{G}_2\mathbf{s}_p(n) + \mathbf{z}_2(n), \qquad n \in \mathcal{N}_{d1}, \tag{8b}$$

respectively. Note that for the CR system, not only the interference from CR to PR, but also that from PR to CR needs to be handled, where the latter case is not considered in [9]. From (8b), we know CR-T1 needs $\mathbf{H}\mathbf{U}_1^*$ and $\mathbf{R}_2$ to determine the optimal transmit covariance matrix for $\mathbf{d}_1(n)$ [12]; and from (8a) we know CR-T1 needs $\mathbf{H}^T\mathbf{U}_2^*$ and $\mathbf{R}_1$ to decode the signal from CR-T2. Similar discussions hold for CR-T2.

If we work on the model (8) directly and train the channel, then CR-T1 can only estimate $\mathbf{H}^T\mathbf{U}_2^*$, while CR-T2 can only estimate $\mathbf{H}\mathbf{U}_1^*$. The knowledge of $\mathbf{H}\mathbf{U}_1^*$ and $\mathbf{R}_2$ have to be fed back from CR-T2 to CR-T1, and the knowledge of $\mathbf{H}^T\mathbf{U}_2^*$ and $\mathbf{R}_1$ have to be fed back from CR-T1 to CR-T2. To release the burden of both channel estimation and feedback,[3] we propose to use *cognitive receive beamforming* at both CR terminals, i.e., CR-T1 and CR-T2 left-multiply the received signals by $\mathbf{U}_1^H$ and $\mathbf{U}_2^H$, and obtain

$$\tilde{\mathbf{y}}_1(n) = \mathbf{U}_1^H\mathbf{H}^T\mathbf{U}_2^*\mathbf{d}_2(n) + \mathbf{U}_1^H\mathbf{z}_1(n) = \mathbf{F}^T\mathbf{d}_2(n) + \tilde{\mathbf{z}}_1(n), \tag{9a}$$

$$\tilde{\mathbf{y}}_2(n) = \mathbf{U}_2^H\mathbf{H}\mathbf{U}_1^*\mathbf{d}_1(n) + \mathbf{U}_2^H\mathbf{z}_2(n) = \mathbf{F}\mathbf{d}_1(n) + \tilde{\mathbf{z}}_2(n), \tag{9b}$$

respectively, where $\mathbf{F}$ and $\tilde{\mathbf{z}}_j(n)$, $j = 1, 2$ represent the equivalent channel and noise, respectively. Some observations are made here:

- The equivalent channels between CRs become reciprocal, which offers advantages as
  - We can estimate the channel at one CR terminal only and then feed it back to the other terminal, which reduces the burden of feedback;
  - We can estimate the channel at both CR terminals and eliminate the necessity of the channel feedback.
- The interference from PR is completely removed at both CR terminals.
- The resultant noise $\tilde{\mathbf{z}}_j(n)$ is still white Gaussian.

[3] Note that, the bandwidth of CR feedback channel is also limited since CR is unlicensed user and could not have much bandwidth.

*2) Practical Case:* With finite learning time, only the estimates $\hat{\mathbf{U}}_j$'s can be obtained. After applying the proposed cognitive beamforming, the two CR terminals receive

$$\mathbf{y}_1(n) = \mathbf{F}^T\mathbf{d}_2(n) + \Delta\mathbf{U}_1^H\mathbf{G}_1\mathbf{s}_p(n) + \tilde{\mathbf{z}}_1(n), \tag{10a}$$

$$\mathbf{y}_2(n) = \mathbf{F}\mathbf{d}_1(n) + \Delta\mathbf{U}_2^H\mathbf{G}_2\mathbf{s}_p(n) + \tilde{\mathbf{z}}_2(n), \tag{10b}$$

where $\mathbf{F}$ and $\tilde{\mathbf{z}}_j(n)$ are now redefined as $\hat{\mathbf{U}}_2^H\mathbf{H}\hat{\mathbf{U}}_1^*$ and $\hat{\mathbf{U}}_1^H\mathbf{z}_1(n)$, respectively.

*Remark 3.1:* With imperfect learning, the channel is still reciprocal and the noise distribution is the same as the perfect learning case. However, there exist residue interferences at the CR receivers caused by PR. Although the interference statistics need to be fed back from one CR terminal to the other for designing the transmit signal covariance, we will later see that in fact only little feedback is needed due to the special structure of $\Delta\mathbf{U}_j^H\mathbf{G}_j\mathbf{s}_p(n)$. Therefore, the advantages in the perfect learning case are mostly maintained even with imperfect learning.

To obtain some essential insights for the optimal design, we will focus on the simplest case in the sequel by setting $\theta = 1$, i.e., transmission only takes place from CR-T1 to CR-T2. The discussion for a general value of $\theta$ can be made based on a similar approach but is rather omitted here for brevity.[4] The covariance matrix of the residue interference $\Delta\mathbf{U}_2^H\mathbf{G}_2\mathbf{s}_p(n)$ can be expressed as

$$\mathbf{E}_2 = \mathrm{E}[\Delta\mathbf{U}_2^H\mathbf{G}_2\mathbf{s}_p(n)\mathbf{s}_p^H(n)\mathbf{G}_2^H\Delta\mathbf{U}_2] = \mathrm{E}[\Delta\mathbf{U}_2^H\mathbf{Q}_2\Delta\mathbf{U}_2] \tag{11}$$

From [13, Eq. (30)] and the fact that $\Delta\mathbf{R}_2 = \Delta\mathbf{R}_2^H$, we know

$$\mathrm{E}[\Delta\mathbf{R}_2\mathbf{\Psi}\Delta\mathbf{R}_2] = \frac{1}{N_l}tr(\mathbf{\Psi}\mathbf{R}_2)\mathbf{R}_2, \tag{12}$$

for any matrix $\mathbf{\Psi}$. Then, we have

$$\begin{aligned}\mathbf{E}_2 =& \mathrm{E}[\mathbf{U}_2^H\Delta\mathbf{R}_2\mathbf{Q}_2^\dagger\mathbf{Q}_2\mathbf{Q}_2^\dagger\Delta\mathbf{R}_2\mathbf{U}_2] = \frac{tr(\mathbf{Q}_2^\dagger\mathbf{R}_2)}{N_l}\mathbf{U}_2^H\mathbf{R}_2\mathbf{U}_2\\ \stackrel{(a)}{=}& \frac{tr(\mathbf{Q}_2^\dagger\mathbf{R}_2)}{N_l}\sigma_{n2}^2\mathbf{I} = \frac{tr(\mathbf{Q}_2^\dagger\mathbf{Q}_2) + \sigma_{n2}^2 tr(\mathbf{Q}_2^\dagger)}{N_l}\sigma_{n2}^2\mathbf{I}\\ =& \frac{\sigma_{n2}^2(M_p + \sigma_{n2}^2 tr(\mathbf{Q}_2^\dagger))}{N_l}\mathbf{I} = \frac{\beta_2}{N_l}\mathbf{I},\end{aligned} \tag{13}$$

[4]Discussing over the general case of $\theta$ requires a more complex mathematical derivations which could, at least, be carried out from the brute-force searching method. However, such an approach would hinder the clear exposition of our learning based cognitive radio scheme and will not to be the focus of this paper.

where "(a)" uses the property that $\mathbf{U}_2^H\mathbf{Q}_2 = \mathbf{0}$, and $\beta_2$ is defined accordingly.

*Remark 3.2:* Interestingly, the interferences at all antennas are uncorrelated and have the same power, $\beta_2$. To assist the source covariance design at CR-T1, only a scalar $\beta_2$ needs to be sent back from CR-T2, which is much easier than feeding back the whole covariance matrix $\hat{\mathbf{R}}_2$. This explains our previous claim in Remark 3.1 that only a little amount of feedback is needed due to the residue interference from PR.

*Remark 3.3:* Computing $\beta_2$ needs some tricks. Since the exact value of $\mathbf{Q}_2$ is not available at CR-T2, we may replace $\mathbf{Q}_2$ by its ML estimate $\hat{\mathbf{Q}}_2$ that can be obtained from $\hat{\mathbf{R}}_2$ according to the algorithms in [9].

Another impact of imperfect channel learning is the CR's residual interference to PR, which is normally characterized by the IT defined as the total interference power at PR [8] expressed as, e.g., for CR-T1,

$$I_{d1} = \mathrm{E}[\|\mathbf{G}_1^T\hat{\mathbf{U}}_1^*\mathbf{d}_1(n)\|^2] = \mathrm{E}[\|\mathbf{G}_1^T\Delta\mathbf{U}_1^*\mathbf{d}_1(n)\|^2]. \tag{14}$$

Although a more accurate characterization should be the performance loss at PR due to the interference [14], such kind of technique requires certain cooperation between the CR and PR. Nonetheless, IT has been proved effective to upper bound the capacity loss at PR [8], [14].

Define $\mathbf{R}_{d1} = \mathrm{E}[\mathbf{d}_1(n)\mathbf{d}_1^H(n)]$ as the transmit covariance matrix of CR-T1. It can be further shown that

$$\begin{aligned} I_{d1} &\stackrel{(a)}{=} \frac{\sigma_{n1}^2 tr(\mathbf{R}_{d1})}{N_l} tr(\mathbf{G}_1^H\mathbf{Q}_1^\dagger\mathbf{R}_1\mathbf{Q}_1^\dagger\mathbf{G}_1) \\ &= \frac{\sigma_{n1}^2 tr(\mathbf{R}_{d1})}{N_l}\Big(tr(\mathbf{G}_1^H\mathbf{Q}_1^\dagger\mathbf{G}_1) + \sigma_{n1}^2 tr(\mathbf{G}_1^H\mathbf{Q}_1^\dagger\mathbf{Q}_1^\dagger\mathbf{G}_1)\Big) \\ &= \frac{tr(\mathbf{R}_{d1})\sigma_{n1}^2(M_p + \sigma_{n1}^2 tr(\mathbf{Q}_1^\dagger))}{\alpha\sigma_s^2 N_l} = \frac{tr(\mathbf{R}_{d1})\beta_1}{\alpha\sigma_s^2 N_l}, \end{aligned} \tag{15}$$

where "(a)" comes from $\mathbf{U}_1^H\mathbf{Q}_1 = \mathbf{0}$, and $\beta_1$ is defined as the corresponding term. An important observation is that the IT is inversely proportional to the learning time $N_l$.

*Example 3.1:* Consider a CR system with parameters $M_p = 2$, $\alpha = 0.5$, $M_1 = M_2 = 4$, $N = 1000$, and $\sigma_{n1}^2 = 1$. We numerically examine the theoretical expression of the IT for $\sigma_s^2 = 0$ dB and $\sigma_s^2 = 20$ dB, respectively. The ML estimate $\hat{\mathbf{Q}}_1$ is used to compute $\beta_1$ for different values of $N_l$. Totally $10,000$ Monte-Carlo runs are taken for averaging. The figure of merit is the inverse of the normalized IT $1/(\sigma_s^2 I_{d1})$. As shown in Fig. 3, the numerical and theoretical results match each other quite well. The higher value of $\sigma_s^2$ yields lower value of IT due to the smaller $\beta_1$.

Suppose the acceptable IT at PR is no more than $\zeta$. Then, the source covariance design at CR-T1 should take care of the following constraint:

$$tr(\mathbf{R}_{d1}) \leq \frac{\zeta\alpha\sigma_s^2 N_l}{\beta_1} = \chi_1 N_l, \tag{16}$$

where $\chi_1$ is defined as $\chi_1 = \zeta\alpha\sigma_s^2/\beta_1$.

*Remark 3.4:* Note that, the parameter $\zeta\alpha\sigma_s^2$ should be obtained by CR via some dedicated means. For example, PR could report to a central controller about this single parameter from time to time, and CR could directly obtain this parameter from the central controller. However, CR does not need to know the instant status of PR as transmitting or receiving.

*Remark 3.5:* From (16), CR-T1 needs to know $\beta_1$ before designing the system parameters, $N_l$, $N_t$, and $N_d$. However, computation of $\beta_1$, similarly as shown in Remark 3.3, depends on $\hat{\mathbf{Q}}_1^\dagger$, which is only available after the learning stage. This looks like a chicken-egg problem. Fortunately, it can be shown that $\beta_1$ varies negligibly when $N_l$ becomes large. From the first-order perturbation analysis in (7), we know $\mathbf{Q}_1 - \hat{\mathbf{Q}}_1$ is of the order $\frac{1}{\sqrt{N_l}}$. Hence, $tr(\hat{\mathbf{Q}}_1^\dagger) = tr(\mathbf{Q}_1^\dagger) + O(\frac{1}{\sqrt{N_l}})$ does not vary much when $N_l$ is large and will finally converge to $tr(\mathbf{Q}_1^\dagger)$. For practical implementation, we may let CR-T1 dynamically learn the channel, and at the same time check whether $\beta_1$ becomes a relatively stable value. Suppose $\beta_1$ is relatively stable when CR learns the channel for $N_0$ symbol periods. Then, CR-T1 can compute the optimized parameter $N_l$ according to the algorithms given in the next section. If the optimal $N_l$ is smaller than $N_0$, then CR-T1 immediately proceeds to the channel training stage; otherwise, CR-T1 will keep on learning for another $N_l - N_0$ symbol periods. Therefore, in the design of the system parameters, there is no harm to treat $\beta_1$ as a known constant factor, which also makes $\chi_1$ a constant value.

*Example 3.2:* We consider the same system setup as Example 3.1 and examine the variation of $\beta_1$ with respect to the learning time $N_l$. Both the theoretical and numerical values of $\beta_1$ are shown in Fig. 4, where the former is obtained from the true matrix $\mathbf{Q}_1$ and the latter is obtained via $\hat{\mathbf{Q}}_1$. It is seen that there always exists some value of $N_0$, beyond which $\beta_1$ becomes a relatively stable value. For example, with PR transmit SNR $\sigma_s^2 = 0$ dB, taking $N_0 = 200$ can guarantee a stable $\beta_1$, while for a higher SNR $\sigma_s^2 = 20$ dB, taking $N_0$ as small as $10$ is sufficient to yield a stable $\beta_1$.

### C. Channel Training Stage

The targets of channel estimation for the CR link in the channel training stage are $\mathbf{F}$ at CR-T2 and $\mathbf{F}^T$ at CR-T1. Thanks to the proposed transmit and receive cognitive beamforming, which yields a pair of reciprocal channels, we may train the channel from both directions and thereby eliminate any feedback, or train the channel from one direction only and then feed back the result from one CR terminal to the other. In this paper, we will adopt the second approach to gain tractable and insightful analysis, whereas considering the first approach does not change the basic principle but complicates the discussions.

Without loss of generality, we assume $M_1 \leq M_2$ and let CR-T1 send the training sequence to CR-T2. To protect PR, the training signal from CR-T1, denoted by $\mathbf{t}_1(n)$, must also be precoded by the matrix $\hat{\mathbf{U}}_1^*$. The received signal at CR-T2, after beamforming, is then given by

$$\tilde{\mathbf{y}}_2(n) = \mathbf{F}\mathbf{t}_1(n) + \Delta\mathbf{U}_2^H\mathbf{G}_2\mathbf{s}_p(n) + \tilde{\mathbf{z}}_2(n), \quad N_l + 1 \leq n \leq N_l + N_t. \tag{17}$$

Denote

$$\begin{aligned}
\tilde{\mathbf{Y}}_2 &= [\tilde{\mathbf{y}}_2(N_l+1), \tilde{\mathbf{y}}_2(N_l+2), \ldots, \tilde{\mathbf{y}}_2(N_l+N_t)] \\
\mathbf{T}_1 &= [\mathbf{t}_1(N_l+1), \mathbf{t}_1(N_l+2), \ldots, \mathbf{t}_1(N_l+N_t)] \\
\mathbf{S}_p &= [\mathbf{s}_p(N_l+1), \mathbf{s}_p(N_l+2), \ldots, \mathbf{s}_p(N_l+N_t)] \\
\tilde{\mathbf{Z}}_2 &= [\tilde{\mathbf{z}}_2(N_l+1), \tilde{\mathbf{z}}_2(N_l+2), \ldots, \tilde{\mathbf{z}}_2(N_l+N_t)].
\end{aligned}$$

The covariance matrix of $\mathbf{F}$ is then computed as as

$$\mathbf{R}_F = \mathrm{E}[\mathbf{F}^H\mathbf{F}] = \mathrm{E}[\hat{\mathbf{U}}_1^T\mathbf{H}^H\hat{\mathbf{U}}_2\hat{\mathbf{U}}_2^H\mathbf{H}\hat{\mathbf{U}}_1^*] = K_2\mathbf{I}. \tag{18}$$

where $K_j \triangleq M_j - M_p$, $j = 1, 2$, for notation simplicity.

The linear-minimum-mean-square-error (LMMSE) -based channel estimator for $\mathbf{F}$ can be obtained as [15]

$$\hat{\mathbf{F}} = \tilde{\mathbf{Y}}_2(\mathbf{T}_1^H\mathbf{R}_F\mathbf{T}_1 + \mathrm{E}[\mathbf{S}_p^H\mathbf{G}_2^H\Delta\mathbf{U}_2\Delta\mathbf{U}_2^H\mathbf{G}_2\mathbf{S}_p] + \sigma_{n2}^2K_2\mathbf{I})^{-1}\mathbf{T}_1^H\mathbf{R}_F, \tag{19}$$

and $\mathrm{E}[\mathbf{S}_p^H\mathbf{G}_2^H\Delta\mathbf{U}_2\Delta\mathbf{U}_2^H\mathbf{G}_2\mathbf{S}_p]$ is separately computed as

$$\mathrm{E}[\mathbf{S}_p^H\mathbf{G}_2^H\Delta\mathbf{U}_2\Delta\mathbf{U}_2^H\mathbf{G}_2\mathbf{S}_p] = \frac{K_2\beta_2}{N_l}\mathbf{I}, \tag{20}$$

where we use the property that $\mathrm{s}_p(n)$'s are temporarily and spatially independent. Substituting (20) and (18) into (19), we obtain

$$\hat{\mathbf{F}} = \tilde{\mathbf{Y}}_2\left(\mathbf{T}_1^H\mathbf{T}_1 + \underbrace{\left(\frac{\beta_2}{N_l} + \sigma_{n2}^2\right)}_{\gamma_2}\mathbf{I}\right)^{-1}\mathbf{T}_1^H, \tag{21}$$

where $\gamma_2$ is defined accordingly. Let $\Delta\mathbf{F} = \mathbf{F} - \hat{\mathbf{F}}$. From the nature of the LMMSE estimation, $\Delta\mathbf{F}$ is uncorrelated with $\hat{\mathbf{F}}$. The rows of $\Delta\mathbf{F}$ are uncorrelated among themselves and each has the covariance

$$\mathbf{R}_{\Delta f} = \left(\mathbf{I} + \frac{1}{\gamma_2}\mathbf{T}_1\mathbf{T}_1^H\right)^{-1}. \tag{22}$$

Moreover, the covariance matrix of each row of $\hat{\mathbf{F}}$ can be calculated as

$$\mathbf{R}_{\hat{f}} = \mathbf{T}_1(\mathbf{T}_1^H\mathbf{T}_1 + \gamma_2\mathbf{I})^{-1}\mathbf{T}_1^H = \mathbf{I} - \mathbf{R}_{\Delta f}. \tag{23}$$

Assuming $\tilde{\mathrm{s}}_p(n)$ to be Gaussian random variables, the entries of $\hat{\mathbf{F}}$ and $\Delta\mathbf{F}$ are easily seen to be Gaussian distributed for a given $\mathbf{G}_2$.

Due to imperfect learning, the residue interference $\mathbf{G}_1^T\Delta\mathbf{U}_1^*\mathbf{t}_1(n)$ is non-zero at PR. The IT caused during training is computed as

$$I_{t1}(n) = \mathrm{E}[\|\mathbf{G}_1^T\hat{\mathbf{U}}_1^*\mathbf{t}_1(n)\|^2] = \frac{\|\mathbf{t}_1(n)\|^2\beta_1}{\alpha\sigma_s^2N_l}. \tag{24}$$

In fact, it is not possible to restrict the instant interference $I_{t1}(n)$ at time slot $n$. Therefore, we will deal with the average interference during the entire training stage, defined as

$$I_{tav} = \frac{1}{N_t}\sum_{n=N_l+1}^{N_l+N_t} I_{1t}(n) = \frac{\beta_1 tr(\mathbf{T}_1\mathbf{T}_1^H)}{\alpha\sigma_s^2N_lN_t}. \tag{25}$$

The IT constraint is then $I_{tav} \leq \zeta$, which is equivalent to

$$tr(\mathbf{T}_1\mathbf{T}_1^H) \leq \chi_1N_lN_t. \tag{26}$$

## IV. CR Transmission Optimization

The tradeoff of power and time allocation between channel training and data transmission has been studied in, e.g., [16] for the traditional multi-antenna system. However for the proposed CR scheme, an additional time period should be assigned for learning. Intuitively, one would expect the larger $N_l$ to get better space estimation such that both the interferences to and from PR can be reduced via cognitive beamforming. However, increasing $N_l$ will decrease $N_d$ for fixed $N_t$ and $N$, and thus reduce the overall system throughput. Meanwhile, the IT constraints during both training and data transmission should be taken into consideration, which bereaves the freedom of the power allocation. All the above issues make the pertinent analysis for the CR system a non-trivial one as compared to the existing results in [16].

Similar to [16], we will evaluate the performance of the proposed CR scheme considering the lower bound on the system ergodic capacity, which is related to both channel estimation errors and residue interferences to and from PR. Based on this lower bound, the optimal power and time allocation over CR's learning, training, and data transmission stages are derived, which provides some insightful guidance for the practical system design. We assume an error-free feedback channel from CR-T2 to CR-T1. The effect of imperfect feedback on the achievable transmission rate has been partly discussed in [17], [18]. As mentioned before, we only focus on the case of $\theta = 1$, i.e., CR-T1 transmitting to CR-T2 in the entire data transmission stage.

Assume the total power that can be allocated to CR-T1 over one frame is $P$, and denote the average powers during training and data transmission over all $M_1$ antennas as $\rho_t$ and $\rho_d$, respectively; namely

$$\mathrm{E}\{\|\hat{\mathbf{U}}_1^*\mathbf{T}_1\|_F^2\} = tr(\mathbf{T}_1\mathbf{T}_1^H) = \rho_t N_t, \qquad \mathrm{E}\{\|\hat{\mathbf{U}}_1^*\mathbf{d}_1(n)\|^2\} = tr(\mathbf{R}_{d1}) = \rho_d, \tag{27}$$

where $\|\cdot\|_F$ denotes the Frobenius norm. Note that the precoding matrix $\mathbf{U}_1^*$ should be taken into account when we compute the power over transmit antennas.

Conservation of time and power yields

$$N = N_l + N_t + N_d, \qquad P \geq \rho_t N_t + \rho_d N_d. \tag{28}$$

Note that "$\geq$" is used in the power allocation constraint to account for the cases when $P$ cannot be fully utilized due to the IT constraints at PR.

### A. Lower Bound on CR Ergodic Capacity

During the data transmission stage, the received signal at CR-T2 can be rewritten as

$$\tilde{\mathbf{y}}_2(n) = \hat{\mathbf{F}}\mathbf{d}_1(n) + \underbrace{\Delta\mathbf{F}\mathbf{d}_1(n) + \Delta\mathbf{U}_2^H\mathbf{G}_2\mathbf{s}_p(n) + \tilde{\mathbf{z}}_2(n)}_{\mathbf{v}_2(n)}, \quad N_l + N_t + 1 \le n \le N, \tag{29}$$

where $\mathbf{v}_2(n)$ is defined as the effective interference-plus-noise term. The covariance matrix of the second term on the right-hand side (RHS) of (29) can be computed as

$$\mathrm{E}[\Delta\mathbf{F}\mathbf{d}_1(n)\mathbf{d}_1^H(n)\Delta\mathbf{F}^H] = tr(\mathbf{R}_{d1}\mathbf{R}_{\Delta f})\mathbf{I}, \tag{30}$$

where the uncorrelation among rows of $\Delta\mathbf{F}$ is utilized. Therefore, $\mathbf{v}_2(n)$ has the covariance

$$\mathbf{R}_{v2} = (tr(\mathbf{R}_{d1}\mathbf{R}_{\Delta f}) + \gamma_2)\mathbf{I}. \tag{31}$$

Note that $\mathbf{v}_2(n)$ is uncorrelated with the signal part $\hat{\mathbf{F}}\mathbf{d}_1(n)$; however, it is not necessarily independent with the signal part.

Since the channel is memoryless, the instantaneous mutual information (IMI) between the unknown data and the observed values at CR-T2 is

$$\begin{aligned}\mathcal{I}(\tilde{\mathbf{y}}_2(n), \tilde{\mathbf{Y}}_2, \mathbf{T}_1; \mathbf{d}_1(n)) \ge& \mathcal{I}(\tilde{\mathbf{y}}_2(n), \hat{\mathbf{F}}; \mathbf{d}_1(n)) \\ =& \mathcal{I}(\tilde{\mathbf{y}}_2(n); \mathbf{d}_1(n)|\hat{\mathbf{F}}) + \underbrace{\mathcal{I}(\hat{\mathbf{F}}; \mathbf{d}_1(n))}_{=0}, \quad N_l + N_t + 1 \le n \le N. \end{aligned} \tag{32}$$

<u>Lemma</u> *4.1:* With instant knowledge of channel $\hat{\mathbf{F}}$, the ergodic capacity of CR channel is lower-bounded by

$$\begin{aligned} C \ge C_{L1} = \max_{\mathbf{T}_1} \mathrm{E}_{\hat{\mathbf{F}}}\left[\max_{\mathbf{R}_{d1}} \log|\mathbf{I} + \mathbf{R}_{v2}^{-1}\hat{\mathbf{F}}\mathbf{R}_{d1}\hat{\mathbf{F}}^H|\right], \\ \text{s.t. } tr(\mathbf{R}_{d1}) = \rho_d \le \chi_1 N_l, \qquad tr(\mathbf{T}_1\mathbf{T}_1^H) = \rho_t N_t \le \chi_1 N_l N_t, \end{aligned} \tag{33}$$

where $\mathrm{E}[\cdot]$ is taken over $\hat{\mathbf{F}}$, and the two constraints are due to the IT constraints (16) and (26), respectively.

*Proof:* See Appendix I. ■

Note that in (33), $\mathbf{R}_{d1}$ should be maximized inside $\mathrm{E}[\cdot]$ because CR-T1 knows $\hat{\mathbf{F}}$ instantaneously. However, training sequence should be fixed for all the channel realizations, and thus $\mathbf{T}_1$ is placed outside $\mathrm{E}[\cdot]$.

### B. Optimizing Training Sequence

Due to the difficulty of computing the optimal $\mathbf{T}_1$ from (33), we will design training sequence based on a different criterion of minimizing the channel estimation mean square error (MSE), i.e., $tr(\mathbf{R}_{\Delta f})$, which is a practically adopted method for channel estimation [19], [20].[5] The similar approach has also been suggested in [16], [21], [22] from different viewpoints. Hence, the optimal training design is found from the problem

$$\begin{aligned} \min_{\mathbf{T}_1} \quad & tr(\mathbf{I} + \frac{1}{\gamma_2}\mathbf{T}_1\mathbf{T}_1^H)^{-1} \\ \text{s.t.} \quad & tr(\mathbf{T}_1\mathbf{T}_1^H) = \rho_t N_t, \end{aligned} \tag{34}$$

where we leave the IT constraint $\rho_t \leq \chi_1 N_l$ in the later optimization. By applying the geometric-arithmetic mean inequality, the optimal $\mathbf{T}_1\mathbf{T}_1^H$ can be easily calculated as $\frac{\rho_t N_t}{K_1}\mathbf{I}$ and the corresponding $\mathbf{R}_{\Delta f}$ is

$$\mathbf{R}_{\Delta f} = \frac{\gamma_2 K_1}{\gamma_2 K_1 + \rho_t N_t}\mathbf{I} = \eta_2 \mathbf{I}. \tag{35}$$

### C. Optimization Over Source Covariance

With the separately designed training, a new lower bound of the ergodic channel capacity is written as

$$C_{L2} = \mathrm{E}_{\hat{\mathbf{F}}}\left[\max_{\mathbf{R}_{d1}} \log\left|\mathbf{I} + \frac{1}{\eta_2 tr(\mathbf{R}_{d1}) + \gamma_2}\hat{\mathbf{F}}\mathbf{R}_{d1}\hat{\mathbf{F}}^H\right|\right], \tag{36}$$

and the constraint is $tr(\mathbf{R}_{d1}) = \rho_d$, where we leave the IT constraint $\rho_d \leq \chi_1 N_l$ in the later optimization.

Define $\hat{\mathbf{F}}^w = \hat{\mathbf{F}}\mathbf{R}_{\hat{f}}^{-1/2} = (1-\eta_2)^{-1/2}\hat{\mathbf{F}}$ as a row-whitened version of $\hat{\mathbf{F}}$. Since the entries of $\hat{\mathbf{F}}^w$ are random Gaussian variables with zero means and unit variances, the distribution of $\hat{\mathbf{F}}^w$ is not related to the system parameters, $\rho_t, \rho_d, N_l, N_t$, and $N_d$. Let the EVD of $(\hat{\mathbf{F}}^w)^H\hat{\mathbf{F}}^w$ be $\mathbf{Q}\mathbf{\Lambda}\mathbf{Q}^H$, where $\mathbf{Q}$ is an unitary matrix and $\mathbf{\Lambda} = \mathrm{diag}\{\lambda_1, \lambda_2, \ldots, \lambda_{K_1}\}$,[6] with $\lambda_i$'s being arranged in a non-increasing order. The distributions of $\lambda_i$'s are not related to the system parameters, either. Define $\mathbf{X} = \frac{(1-\eta_2)}{\eta_2\rho_d+\gamma_2}\mathbf{Q}^H\mathbf{R}_{d1}\mathbf{Q}$. Then, the capacity lower bound is rewritten as

$$C_{L2} = \mathrm{E}_{\lambda_i}\left[\max_{\substack{\mathbf{X}:tr(\mathbf{X})=\rho_{\text{eff}}\\ \mathbf{X}\succcurlyeq\mathbf{0}}} \log\left|\mathbf{I}+\rho_{\text{eff}}\mathbf{X}\mathbf{\Lambda}\right|\right], \tag{37}$$

where

$$\rho_{\text{eff}} = \frac{\rho_d(1-\eta_2)}{\eta_2\rho_d+\gamma_2} = \frac{\rho_d\rho_t N_t}{\gamma_2(\rho_d K_1+\gamma_2 K_1+\rho_t N_t)} \tag{38}$$

is defined as the effective signal-to-noise ratio (SNR). It is easily known that the optimal $\mathbf{X}$ possesses a diagonal structure $\mathbf{X} = \mathrm{diag}\{x_1, x_2, \ldots, x_{K_1}\}$, whose value is found from the standard water-filling algorithm [25] as

$$x_i = \left(\mu - \frac{1}{\lambda_i}\right)^+, \tag{39}$$

where $(\cdot)^+$ denotes $\max(\cdot, 0)$, and $\mu$ represents the water-level chosen to satisfy $\sum_{i=1}^{K_1} x_i = \rho_{\text{eff}}$.

[5]Considering MSE-based channel estimation does not deteriorate the main merit of the proposed study since we aim to provide a practical design.

[6]Recall that we have assumed that $K_1 \leq K_2$.

Define $q_k = \frac{k}{\lambda_{k+1}} - \sum_{j=1}^{k}\frac{1}{\lambda_j}$ for $k = 1, \ldots, K_1 - 1$, $q_0 = 0$, and $q_{K_1} = +\infty$. Then, $C_{L2}$ is expressed as $C_{L2} = \mathrm{E}_{\lambda_i}[g(\rho_{\text{eff}}, \lambda_i)]$, where $g(\rho_{\text{eff}}, \lambda_i)$ is a segment function:

$$g(\rho_{\text{eff}}, \lambda_i) = \sum_{i=1}^{k}\log\left(\frac{\lambda_i}{k}\left(\rho_{\text{eff}} + \sum_{j=1}^{k}\frac{1}{\lambda_j}\right)\right), \qquad \rho_{\text{eff}} \in (q_{k-1}, q_k]. \tag{40}$$

*Lemma 4.2:* For given $\lambda_i$'s, $g(\rho_{\text{eff}}, \lambda_i)$ is a continuous, differentiable, increasing, and concave function of $\rho_{\text{eff}}$.

*Proof:* See [9]. ■

*Corollary 4.1:* $C_{L2}$ is a continuous, differentiable, increasing, and concave function of $\rho_{\text{eff}}$.

*Proof:* Apply Lemma 4.2 and the property that the distributions of $\lambda_i$'s are independent from $\rho_{\text{eff}}$. ■

### D. Optimization Over Power Allocation

Averaged over the entire CR frame, the lower bound on the ergodic capacity becomes

$$C_{AL} = \frac{N_d}{N}C_{L2}, \tag{41}$$

where $N_d/N$ accounts for the fact that the data transmission occupies an interval of $N_d$ symbols. $C_{AL}$ in (41) is a function of different system parameters, $\rho_d, \rho_t, N_l, N_t$, and $N_d$, whose optimal values should be obtained by maximizing $C_{AL}$. From now on, we will virtually consider $N_l$, $N_t$, and $N_d$ as continuous variables. It needs to be mentioned that $N_t$ should be no less than $K_1$ in order to obtain a meaningful channel estimation, while $N_d$ must be no less than $1$ in order to achieve a meaningful transmission.

Since $C_{L2}$ is an increasing function of $\rho_{\text{eff}}$, we can equivalently obtain the optimal $\rho_t$ and $\rho_d$ by maximizing $\rho_{\text{eff}}$. Considering the IT constraints and the total power constraint, this optimization problem is expressed as

$$\max_{\rho_t,\rho_d} \quad \rho_{\text{eff}} \tag{42a}$$

$$\text{s.t.} \quad \rho_t \leq \chi_1 N_l, \tag{42b}$$

$$\rho_d \leq \chi_1 N_l, \tag{42c}$$

$$\rho_t N_t + \rho_d N_d \leq P, \tag{42d}$$

for given $N_l$ and $N_t$. The optimization problem (42) is non-convex, while we will, in the following, derive its closed-form solutions.

By carefully observing the above three constraints, we find that if $\chi_1 N_l(N - N_l) \leq P$, then (42b) and (42c) hold with equalities for the optimal solution, because $\rho_{\text{eff}}$ is an increasing function of both $\rho_d$ and $\rho_t$. Otherwise, the equality in (42d) must hold. Define $\mathcal{T}_l$ as the set of $N_l$ with $\chi_1 N_l(N - N_l) \leq P$ (the explicit expression of $\mathcal{T}_l$ is omitted here). Obviously, $\mathcal{T}_l$ is a constant set that can be computed before the optimization. Based on the above discussion, we consider the following two cases:

*Case 1: $N_l \in \mathcal{T}_l$ :* In this case, the optimal power allocation is $\rho_t^\star = \rho_d^\star = \chi_1 N_l$. The effective SNR is

$$\text{S1):} \qquad \rho_{\text{eff}}^\star = \frac{\chi_1^2 N_l^2 N_t}{\gamma_2(\chi_1 N_l K_1 + \gamma_2 K_1 + \chi_1 N_l N_t)}. \tag{43}$$

*Case 2: $N_l \notin \mathcal{T}_l$:* In this case, $\rho_{\text{eff}}$ becomes

$$\rho_{\text{eff}} = \frac{\rho_d(P - \rho_d N_d)}{\gamma_2(P + \gamma_2 K_1 - \rho_d(N_d - K_1))} = \frac{(P - \rho_t N_t)\rho_t N_t}{\gamma_2((P - \rho_t N_t)K_1 + \gamma_2 K_1 N_d + \rho_t N_t)}. \tag{44}$$

To proceed, we first ignore the constraints (42b) and (42c), and denote the solutions that maximize (44) as $\rho_d'$ and $\rho_t'$, respectively. Define $c = \frac{(P+\gamma_2 K_1)N_d}{(N_d - K_1)P}$ for $N_d \neq K_1$. Following the similar approach in [16], we know (44) has only one valid root of $\rho_d'$ in the region $[0, P/N_d]$ (or one valid root of $\rho_t'$ in the region $[0, P/N_t]$), and the solutions are expressed as

$$\rho_d' = \frac{P}{N_d} \times \begin{cases} c - \sqrt{c(c-1)} & N_d > K_1 \\ \frac{1}{2} & N_d = K_1 \\ c + \sqrt{c(c-1)} & N_d < K_1 \end{cases}, \tag{45}$$

$$\rho_t' = \frac{P - \rho_d' N_d}{N_t}. \tag{46}$$

The corresponding effective SNR is obtained as

$$\rho_{\text{eff}}' = \begin{cases} \frac{P}{\gamma_2(N_d - K_1)}(\sqrt{c} - \sqrt{c-1})^2 & N_d > K_1 \\ \frac{P^2}{4\gamma_2 K_1 (P + \gamma_2 K_1)} & N_d = K_1 \\ \frac{P}{\gamma_2(K_1 - N_d)}(\sqrt{-c} - \sqrt{1-c})^2 & N_d < K_1 \end{cases}. \tag{47}$$

The following lemma is very important for the later discussions.

*Lemma 4.3:* For a given $N_l$, $\rho_d'$ is an increasing (decreasing) function of $N_t$ ($N_d$), while $\rho_t'$ is a decreasing (increasing) function of $N_t$ ($N_d$).

*Proof:* See Appendix II. ■

Define $P_t' = \rho_t' N_t$ and $P_d' = \rho_d' N_d$ as the corresponding powers allocated to training and data transmission.

*Lemma 4.4:* For a given $N_l$, $P_d'$ is a decreasing (increasing) function of $N_t$ ($N_d$), while $P_t'$ is a decreasing (increasing) function of $N_t$ ($N_d$).

*Proof:* The proof follows the similar method given in Appendix II. ■

Now let us include back the constraints (42b) and (42c) to derive the true optimal solutions of (42). There exist the following three subcases:

1) $\rho_t' \geq \chi_1 N_l$: Since (44) has only one valid root $\rho_t'$, the optimal $\rho_t$ considering (42b) must stay on the boundary, which gives $\rho_t^\star = \chi_1 N_l$. Then, the optimal $\rho_d$ is directly computed as $\rho_d^\star = (P - \chi_1 N_l N_t)/N_d$. The corresponding effective SNR is

$$\text{S2):} \qquad \rho_{\text{eff}}^\star = \frac{(P - \chi_1 N_l N_t)\chi_1 N_l N_t}{\gamma_2((P - \chi_1 N_l N_t)K_1 + \gamma_2 K_1 N_d + \chi_1 N_l N_t N_d)}. \tag{48}$$

Since $\rho_t'$ is a decreasing function of $N_t$, the region of $N_t$ for this subcase can be represented by $\mathcal{T}_{t1}(N_l) = [K_1, N_1]$, where $N_1$ can be computed from (45) as the value of $N_t$ that makes $\rho_t' = \chi_1 N_l$.

2) $\rho'_d \geq \chi_1 N_l$: Similar to the previous subcase, we obtain $\rho_d^\star = \chi_1 N_l$ and $\rho_t^\star = (P - \chi_1 N_l N_d)/N_t$. The optimal effective SNR is

$$\text{S3):} \qquad \rho_{\text{eff}}^\star = \frac{(P - \chi_1 N_l N_d)\chi_1 N_l}{\gamma_2(\chi_1 N_l K_1 + \gamma_2 K_1 + P - \chi_1 N_l N_d)}. \tag{49}$$

Since $\rho'_d$ is a decreasing function of $N_d$, the region of $N_d$ for this subcase can be represented by $\mathcal{T}_d(N_l) = [1, N_2]$, where $N_2$ can be computed from (45) as the value of $N_d$ that makes $\rho'_d = \chi_1 N_l$. Correspondingly, the range of $N_t$ in this subcase is denoted by $\mathcal{T}_{t2}(N_l) = [N - N_l - N_2, N - N_l - 1]$.

3) Otherwise, $\rho_t^\star = \rho'_t$, $\rho_d^\star = \rho'_d$, and neither (42b) nor (42c) holds in equality. The range of $N_t$ in this subcase is immediately obtained as $\mathcal{T}_{t3}(N_l) = [N_1, N - N_l - N_2]$, and the corresponding effective SNR is

$$\text{S4):} \qquad \rho_{\text{eff}}^\star = \rho'_{\text{eff}}. \tag{50}$$

Fig. 5 is quite helpful for understanding where the subcases S2), S3), and S4) take place.

*Example 4.1:* The same system setup as Example 3.1 is used here. Two new parameters are introduced as $P = 20,000$ and $\chi_1 = 0.16$. The optimal $\rho_t^\star$ and $\rho_d^\star$ versus $N_t$ at $N_l = 200$ are shown in Fig. 6. The following observations are made:

- $\rho_t^\star$ is constant over S2) since $\rho'_d$ is bounded by $\chi_1 N_l$; $\rho_t^\star$ is decreasing over S4) since it is equivalent to $\rho'_t$, which is a decreasing function of $N_t$ from Lemma 4.3; $\rho_t^\star$ is increasing over S3) as is seen from $\rho_t^\star = (P - \chi_1 N_l N_d)/N_t$;
- $\rho_d^\star$ is decreasing over S2) as is seen from $\rho_d^\star = (P - \chi_1 N_l N_t)/N_d$; $\rho_d^\star$ is increasing over S4) since it is equivalent to $\rho'_d$; $\rho_d^\star$ is constant over S3) since it is bounded by $\chi_1 N_l$.

Define $P_t^\star = \rho_t^\star N_t$ and $P_d^\star = \rho_d^\star N_d$ as the powers allocated to training and data transmission. We plot $P_t^\star$ and $P_d^\star$ versus $N_t$ in Fig. 7. From the observations in Fig. 6, we know that for subcases S2) and S3), $P_d^\star$ is a decreasing function of $N_t$, while $P_t^\star$ is an increasing function of $N_t$. Furthermore, since $P_t^\star = P'_t$ and $P_d^\star = P'_d$ over S4), from Lemma 4.4 we know that the increasing property of $P_t^\star$ and the decreasing property of $P_d^\star$ are kept over S4), too.

### E. Optimization Over Time Allocation

Substituting the closed-form expression of $\rho_{\text{eff}}$ back to $C_{AL}$, we then formulate the optimization over the remaining variables, $N_l$ and $N_t$, as

$$\max_{N_l, N_t} \quad C_{AL} \tag{51}$$
$$\text{s.t.} \quad N_t \geq K_1, \quad N_d = N - N_l - N_t \geq 1.$$

The discussion is divided into four parts, corresponding to four subcases S1) to S4) in the previous subsection:

*Subcase* S1): From (43), it can be readily checked that $\frac{\partial \rho_{\text{eff}}^\star}{\partial N_t} > 0$, and $\frac{\partial^2 \rho_{\text{eff}}^\star}{\partial^2 N_t} < 0$, for a given $N_l$. Since $C_{L2}$ is an increasing concave function of $\rho_{\text{eff}}^\star$, there is

$$\frac{\partial C_{L2}}{\partial N_t} = \frac{\partial C_{L2}}{\partial \rho_{\text{eff}}^\star} \frac{\partial \rho_{\text{eff}}^\star}{\partial N_t} > 0, \tag{52}$$

$$\frac{\partial^2 C_{L2}}{\partial^2 N_t} = \frac{\partial^2 C_{L2}}{\partial^2 \rho_{\text{eff}}^\star} \left( \frac{\partial \rho_{\text{eff}}^\star}{\partial N_t} \right)^2 + \frac{\partial C_{L2}}{\partial \rho_{\text{eff}}^\star} \frac{\partial^2 \rho_{\text{eff}}^\star}{\partial^2 N_t} < 0. \tag{53}$$

Therefore, $C_{L2}$ is an increasing concave function of $N_t$. Since $\frac{N-N_l-N_t}{N}$ is a linearly decreasing function in $N_t$, by chain rule we know $C_{AL}$ is concave in $N_t$. Therefore, for a given $N_l \in \mathcal{T}_l$, the efficient convex optimization tools can be applied to find $N_t$.

*Subcase* S2): For this subcase, there is no direct clue so we propose a one dimensional search over $N_t \in \mathcal{T}_{t1}(N_l)$.

*Subcase* S4): We provide the following lemma for this subcase:

*Lemma 4.5:* $C_{AL}$ is a decreasing (increasing) function of $N_t$ ($N_d$) over the region $N_t \in \mathcal{T}_{t3}(N_l)$.

*Proof:* See Appendix III. ■

Therefore, we should reduce $N_t$ as much as possible if subcase S4) takes place. So the optimal $N_t$ in this case is simply $N_1$.

*Subcase* S3): We provide the following lemma for this subcase:

*Lemma 4.6:* $C_{AL}$ over $N_t \in \mathcal{T}_{t2}(N_l)$ is smaller than that over $N_t \in \mathcal{T}_{t3}(N_l)$.

*Proof:* Consider the optimization over $N_t \in \mathcal{T}_{t2}(N_l)$ but without the IT constraint (42b) and (42c). Then, $\rho_d'$ and $\rho_t'$ become the optimal power values over $N_t \in \mathcal{T}_{t2}(N_l)$. Similarly as subcase

S4), the resultant optimal capacity lower bound, denoted as $C'_{AL}$, is a decreasing function of $N_t$. Since region $\mathcal{T}_{t3}(N_l)$ is on the right side of $\mathcal{T}_{t2}(N_l)$, as shown in Fig. 5, we know $C'_{AL}$ over $\mathcal{T}_{t3}(N_l)$ is smaller than $C_{AL}$ over $\mathcal{T}_{t2}(N_l)$. Adding the interference constraint back, we know the true optimal $C_{AL}$ over $\mathcal{T}_{t2}(N_l)$ must be smaller than $C'_{AL}$, which must also be smaller than $C_{AL}$ over $\mathcal{T}_{t3}(N_l)$. ■

Based on the above discussions, the optimal time allocation is found from the following rules:

- One dimensional searching of $N_l$ is applied.
  - For any $N_l \in \mathcal{T}_l$, $N_t$ can be efficiently found from the convex optimization tools.
  - For any $N_l \notin \mathcal{T}_l$, only $N_t$ in region S2) needs to be checked. In fact, as shown in Fig. 6, the set $\mathcal{T}_{t1}(N_l)$ is usually of small size.

## V. Simulation Results

In this section, we numerically examine the proposed study using various examples. The system as well as the parameters are the same as those in Example 4.1. We assume that the transmit power of PR is $\sigma_s^2 = 20$ dB, so $N_0 = 10$ can guarantee a very good estimate of $\beta_j$, $j = 1, 2$.

1) *$C_{AL}$ as a function of $N_l$ and $N_t$:* In the first example, we take $\chi_1 = 0.16$ and plot $C_{AL}$ as a function of $N_l$ and $N_t$ in Fig. 8. It is seen that the shape of $C_{AL}$ looks like a tent over the three-dimensional space, and there is a unique peak, where $C_{AL}$ is maximized. Then, we have the following conjecture that remains to be proved.

*Conjecture 1:* $C_{AL}$ is a joint concave function of $N_l$ and $N_t$.

*2) Optimal $N_l$ and $N_t$ as a function of $\chi_1$:* Besides introducing one more parameter $N_l$, the effect of IT is another difference between our proposed work and that in [16]. In this sense, it is of interest to take a look at how the optimal time allocation is affected by the IT requirement. The values of optimal $N_l$ and $N_t$, denoted as $N_l^\star$ and $N_t^\star$ respectively, versus $\chi_1$ are then shown in Fig. 9. We have the following observations:

- $N_l^\star$ is a decreasing function of $\chi_1$. This is because that when higher IT can be tolerated at PR, less learning time could be used to save the learning overhead.
- $N_t^\star$ increases first and then decreases with the increasing of $\chi_l$. The reason is that the optimal $N_t^\star$ is not only a function of $\chi_1$ but is also affected by $N_l$. When $\chi_1$ is small, $\rho_t$ is likely to be bounded by $\chi_1 N_l$ as is seen from (42b). Therefore, the total training power $\rho_t N_t$ may not be sufficient for a small $N_t$, so we have to increase the training time. However, once $\chi_1$ gets larger, sufficient training power can be obtained from a very short training time so $N_t$ should be decreased to save the training overhead for data transmission. Finally, as is also seen in Fig. 9, $N_t$ reduces to its lower bound $K_1 = 2$ when $\chi_1 = 10$.

*3) The maximum $C_{AL}$ as a function of $\chi_1$:* In this example, we would like to take a look at how the maximum capacity lower bound, denoted as $C^\star_{AL}$, varies with different IT power levels $\zeta$. Since $\chi_1 = \frac{\zeta\alpha\sigma_s^2}{\beta_1}$ is a linear scaling of $\zeta$, we instead examine $\chi_1$ and the curve of $C^\star_{AL}$ versus $\chi_1$ is displayed in Fig. 10. To illustrate the effect of the optimal power allocation on the capacity bound, we also consider a new scenario where equal power allocation $\rho_d = \rho_t = \min\{\chi_1 N_l, P/(N - N_l)\}$ is adopted, and the corresponding optimal $C_{AL}$ is obtained by searching all the candidates of $N_l$ and $N_t$. It is first seen that $C^\star_{AL}$ is a non-decreasing function of $\chi_1$, which is intuitively correct. However, when $\chi_1$ is too large, the IT constraints do not take any effect and the capacity bound $C^\star_{AL}$ cannot be increased anymore. Moreover, the equal power allocation provides comparable capacity value as that of the optimal power allocation when $\chi_1$ is small. This is because that at lower $\chi_1$, the optimal power allocation is roughly bounded by the IT as $\rho_d = \rho_t = \chi_1 N_l$, which is the same as the equal power allocation. However, when $\chi_1$ is relatively larger, the equal power allocation becomes suboptimal.

Since the equal-power allocation between training and data transmission can yield relatively good performances, we then demonstrate with this power allocation scheme the achievable rate of a practical modulation and coding scheme (MCS) with the discrete bit granularity $\Delta > 0$. The well-known SNR "gap" approximation, denoted by $\Gamma$, is adopted, which measures the power required by the considered MCS in addition to the minimum power obtained from the standard capacity function to support a given decoding error probability [23]. Then, the optimal discrete bit loading algorithm [24] can be applied to obtain the achievable rate. For a practical MCS with $\Delta = 0.5$ and $\Gamma = 3$ dB, the corresponding achievable rate is also included in Fig. 10, which demonstrates the usefulness of the proposed studies for the practical system design.

## VI. Conclusion

In this work, we studied the design of transmission for a multi-antenna CR link under spectrum sharing with a PR link. Our studies built up two major contributions. First, we proposed a concrete CR deployment strategy that consists of environment learning, channel training, and data transmission stages, where detailed formulations on these stages were provided. Second, by analyzing the system parameters, we developed the algorithms to find the optimal power and time allocation for different stages so as to maximize the lower bound on the CR ergodic capacity. Closed-form solution of power allocation was found for a given time allocation, while the optimal time allocation was found via a two-dimensional searching over a confined set.

## Appendix I

## Proof of Lemma 4.1

We drop the index $n$ here for brevity. The IMI between the output $\tilde{\mathbf{y}}_2$ and $\mathbf{d}_1$ conditioned on the channel estimate $\hat{\mathbf{F}}$ is

$$\mathcal{I}(\tilde{\mathbf{y}}_2;\mathbf{d}_1|\hat{\mathbf{F}}) = h(\mathbf{d}_1|\hat{\mathbf{F}}) - h(\mathbf{d}_1|\tilde{\mathbf{y}}_2,\hat{\mathbf{F}}) = h(\mathbf{d}_1) - h(\mathbf{d}_1|\tilde{\mathbf{y}}_2,\hat{\mathbf{F}}). \tag{54}$$

A lower bound on the capacity is obtained by directly taking $\mathbf{d}_1$ as a Gaussian random vector. In this case, the differential entropy $h(\mathbf{d}_1) = \log(|\pi e \mathbf{R}_{d1}|)$. By definition

$$h(\mathbf{d}_1|\hat{\mathbf{F}},\tilde{\mathbf{y}}_2) = h(\mathbf{d}_1 - f(\tilde{\mathbf{y}}_2)|\hat{\mathbf{F}},\tilde{\mathbf{y}}_2), \tag{55}$$

for any function $f(\cdot)$. Moreover, there is

$$h(\mathbf{d}_1 - f(\tilde{\mathbf{y}}_2)|\hat{\mathbf{F}},\tilde{\mathbf{y}}_2) \le h(\mathbf{d}_1 - f(\tilde{\mathbf{y}}_2)|\hat{\mathbf{F}},\tilde{\mathbf{y}}_2) \le \log(|\pi e \mathrm{Cov}(\mathbf{d}_1 - f(\tilde{\mathbf{y}}_2)|\hat{\mathbf{F}},\tilde{\mathbf{y}}_2)|), \tag{56}$$

where $\mathrm{Cov}(\cdot)$ denotes the covariance matrix of a random vector. To achieve the tightest bound, we wish to find a function $f(\cdot)$, such that $|\mathrm{Cov}(\mathbf{d}_1 - f(\tilde{\mathbf{y}}_2)|\hat{\mathbf{F}},\tilde{\mathbf{y}}_2)|$ is minimized. Since it is hard to find such a function $f(\cdot)$, we will, instead, accept a linear function $f(\tilde{\mathbf{y}}_2) = \mathbf{A}\tilde{\mathbf{y}}_2$ with which $tr(\mathrm{Cov}(\mathbf{d}_1 - \mathbf{A}\tilde{\mathbf{y}}_2|\hat{\mathbf{F}},\tilde{\mathbf{y}}_2))$ is minimized. Therefore, $\mathbf{A}$ is the LMMSE estimator of $\mathbf{d}_1$, given $\hat{\mathbf{F}}$ and $\tilde{\mathbf{y}}_2$, i.e.,

$$\mathbf{A} = \mathbf{R}_{d1}\hat{\mathbf{F}}^H(\hat{\mathbf{F}}\mathbf{R}_{d1}\hat{\mathbf{F}}^H + \mathbf{R}_{v2})^{-1}, \tag{57}$$

where the property that $\hat{\mathbf{F}}$ and $\mathbf{v}_2$ are uncorrelated is used thanks to the LMMSE estimation of $\mathbf{F}$. Therefore,

$$\mathrm{Cov}(\mathbf{d}_1 - \mathbf{A}\tilde{\mathbf{y}}_2|\hat{\mathbf{F}}) = (\mathbf{R}_{d1}^{-1} + \hat{\mathbf{F}}^H \mathbf{R}_{v2}^{-1}\hat{\mathbf{F}})^{-1}, \tag{58}$$

and a lower bound on the capacity is obtained as

$$\mathcal{I}(\tilde{\mathbf{y}}_2; \mathbf{d}_1|\hat{\mathbf{F}}) \geq \log|\mathbf{R}_{d1}(\mathbf{R}_{d1}^{-1} + \hat{\mathbf{F}}^H\mathbf{R}_{v2}^{-1}\hat{\mathbf{F}})| = \log(|\mathbf{I} + \mathbf{R}_{v2}^{-1}\hat{\mathbf{F}}\mathbf{R}_{d1}\hat{\mathbf{F}}^H|). \tag{59}$$

This lower bound is achieved when the input signal $\mathbf{d}_1$ is Gaussian and the effective noise $\mathbf{v}_2$ behaves as Gaussian. Taking the expectation over (59) yields the lower bound on the ergodic capacity. Meanwhile, considering that the variables to be adjusted to maximize this lower bound are $\mathbf{R}_{d1}$ and $\mathbf{T}_1$, Lemma 4.1 thus follows.

## Appendix II
## Proof of Lemma 4.3

We will prove that $\rho_d'$ is a decreasing function of $N_d$ for $N_d > K_1$ and omit the proofs for the other cases since they are quite straightforward. Define $\tilde{c} = \frac{N_d}{N_d - K_1}$ and $\tilde{\alpha} = \frac{P}{P + \gamma_2 K_1}$. It suffices to prove that

$$\Xi \triangleq \frac{1}{N_d}(\tilde{c} - \sqrt{\tilde{c}(\tilde{c} - \tilde{\alpha})}) \tag{60}$$

is a decreasing function of $N_d$. Bearing in mind the following properties:

$$0 \leq \tilde{\alpha} \leq 1, \qquad \frac{\partial \tilde{c}}{\partial N_d} = \frac{-K_1}{(N_d - K_1)^2},$$

we obtain

$$\begin{aligned}
\frac{\partial \Xi}{\partial N_d} &= -\frac{1}{N_d^2}\left(\tilde{c} - \sqrt{\tilde{c}(\tilde{c}-\tilde{\alpha})} + \left(\tilde{c} - \frac{\tilde{c}(\tilde{c} - \tilde{\alpha}/2)}{\sqrt{\tilde{c}(\tilde{c}-\tilde{\alpha})}}\right)\frac{K_1}{N_d - K_1}\right) \\
&= -\frac{1}{N_d^2}\left(\tilde{c} - \sqrt{\tilde{c}(\tilde{c}-\tilde{\alpha})} + \left(\tilde{c} - \sqrt{\tilde{c}(\tilde{c}-\tilde{\alpha})}\right)\frac{K_1}{N_d - K_1} + \frac{\alpha}{2}\sqrt{\frac{\tilde{c}}{\tilde{c}-\tilde{\alpha}}}\frac{K_1}{N_d - K_1}\right) \\
&= -\frac{1}{N_d^2}\left(\left(\tilde{c} - \sqrt{\tilde{c}(\tilde{c}-\tilde{\alpha})}\right)\frac{N_d}{N_d - K_1} - \frac{\alpha}{2}\sqrt{\frac{\tilde{c}}{\tilde{c}-\tilde{\alpha}}}\frac{K_1}{N_d - K_1}\right).
\end{aligned} \tag{61}$$

Since $N_d > K_1$, we only need to prove

$$\tilde{c}N_d > \frac{\alpha K_1}{2}\sqrt{\frac{\tilde{c}}{\tilde{c}-\tilde{\alpha}}} + \sqrt{\tilde{c}(\tilde{c}-\tilde{\alpha})}N_d \tag{62}$$

or equivalently

$$N_d > \frac{K_1}{4(\frac{1-\tilde{\alpha}}{\alpha}\frac{N_d}{K_1}+1)}, \tag{63}$$

which is quite obvious since $\frac{1-\tilde{\alpha}}{\alpha}\frac{N_d}{K_1} > 0$.

## APPENDIX III
## PROOF OF LEMMA 4.5

We will only examine the case $N_d > K_1$, whereas other cases can be handled similarly. First, differentiating $\rho_{\text{eff}}^\star$ with respect to $N_d$ gives

$$\frac{\partial \rho_{\text{eff}}^\star}{\partial N_d} = \frac{P(\sqrt{c}-\sqrt{c-1})^2}{\gamma_2(N_d-K_1)^2}\left(\frac{K_1\sqrt{c}}{N_d\sqrt{c-1}}-1\right) = \frac{\rho_{\text{eff}}^\star}{N_d-K_1}\left(\sqrt{\frac{(P+\gamma_2K_1)K_1}{(P+\gamma_2N_d)N_d}}-1\right). \tag{64}$$

From (33), we only need to prove that $\Omega \triangleq \frac{N_d}{N}g(\rho_{\text{eff}},\lambda_i)$ is an increasing function of $N_d$. The differentiation of $\Omega$ with respect to $N_d$ is given by the segment function

$$\frac{\partial \Omega}{\partial N_d} = \frac{1}{N}\left(\sum_{i=1}^{k} g(\rho_{\text{eff}}^\star,\lambda_i) - \frac{k\rho_{\text{eff}}^\star}{\left(\rho_{\text{eff}}^\star+\sum_{j=1}^{k}\frac{1}{\lambda_j}\right)}\frac{N_d}{N_d-K_1}\left(1-\sqrt{\frac{(P+\gamma_2K_1)K_1}{(P+\gamma_2N_d)N_d}}\right)\right), \tag{65}$$

$$\rho_{\text{eff}}^\star \in (q_{k-1}, q_k].$$

Since $N_d > K_1$, there is

$$\frac{N_d}{N_d-K_1}\left(1-\sqrt{\frac{(P+\gamma_2K_1)K_1}{(P+\gamma_2N_d)N_d}}\right) < 1. \tag{66}$$

It needs to prove that the segment function

$$w(\rho_{\text{eff}}^\star) = \sum_{i=1}^{k} g(\rho_{\text{eff}}^\star,\lambda_i) - \frac{k\rho_{\text{eff}}^\star}{\left(\rho_{\text{eff}}^\star+\sum_{j=1}^{k}\frac{1}{\lambda_j}\right)}, \quad \rho_{\text{eff}}^\star \in (q_{k-1}, q_k] \tag{67}$$

is nonnegative for $\rho_{\text{eff}} \geq 0$.

For the $k$th segment, i.e., $\rho_{\text{eff}}^\star \in [q_{k-1}, q_k]$, there is

$$\lim_{\rho_{\text{eff}}^\star \to q_{k-1}^+} w(\rho_{\text{eff}}^\star) = \sum_{i=1}^{k-1}\log\frac{\lambda_i}{\lambda_k} - \left((k-1)-\sum_{j=1}^{k-1}\frac{\lambda_k}{\lambda_j}\right) = \sum_{i=1}^{k-1}\left(\log\frac{\lambda_i}{\lambda_k}-\frac{\lambda_i-\lambda_k}{\lambda_i}\right). \tag{68}$$

By letting $x = \frac{\lambda_i-\lambda_k}{\lambda_k}$ and using the inequality $\log(1+x)-\frac{x}{1+x} \geq 0$ for $x \geq 0$, we know $w(q_{k-1}) \geq 0$. The differentiation of $w(\rho_{\text{eff}}^\star)$ in the $k$th segment is

$$\frac{\partial w(\rho_{\text{eff}}^\star)}{\partial \rho_{\text{eff}}^\star} = \frac{k\rho_{\text{eff}}^\star}{\left(\rho_{\text{eff}}^\star+\sum_{j=1}^{k}\frac{1}{\lambda_j}\right)^2} \geq 0. \tag{69}$$

Therefore, $\Omega$ is an increasing function of $N_d$ over all segments.

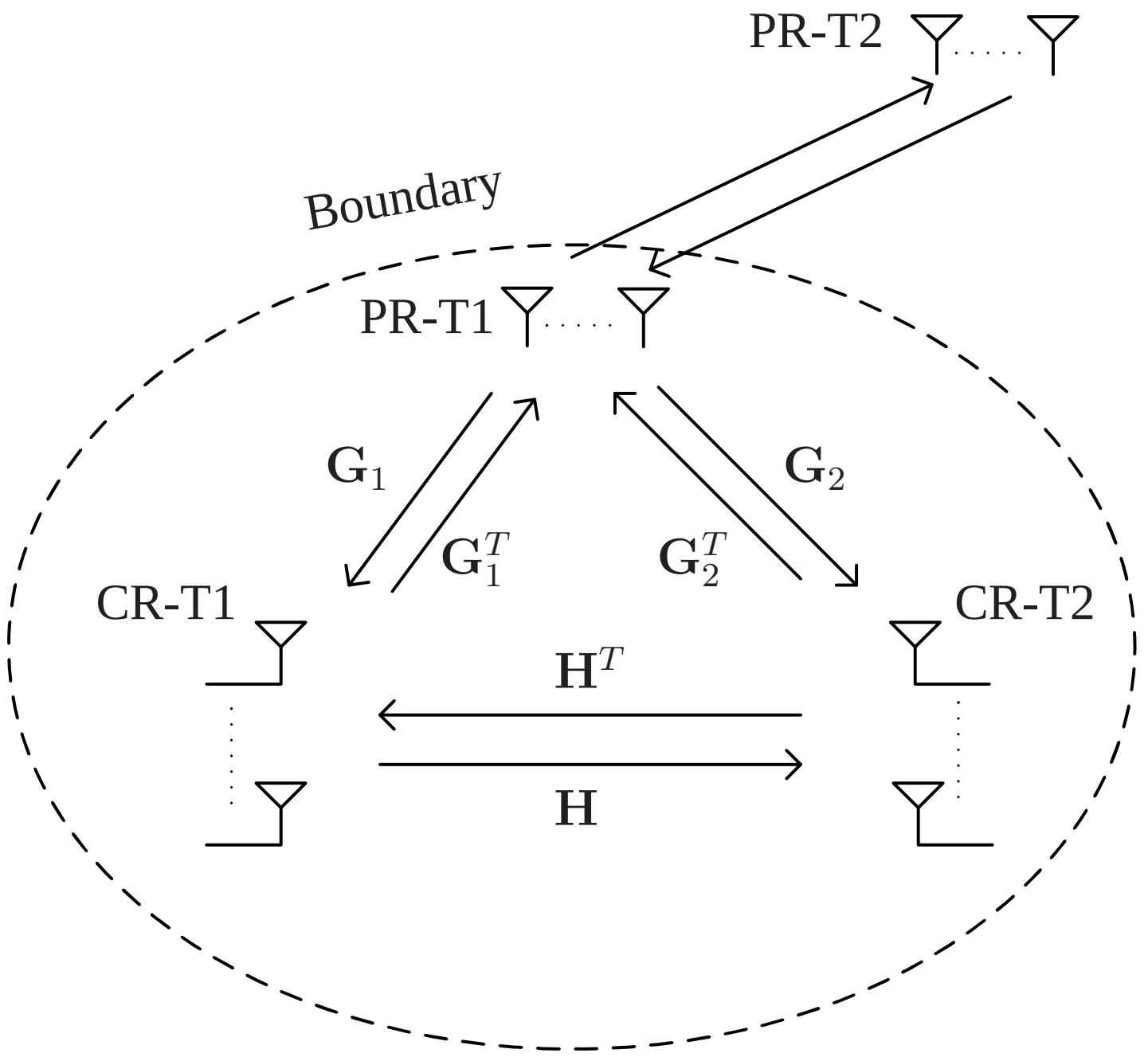

Fig. 1. System model for the multi-antenna CR system.

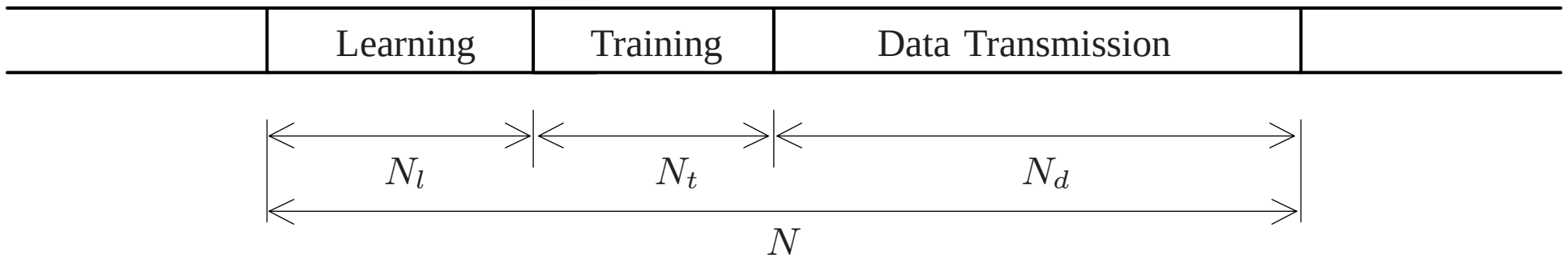

Fig. 2. CR frame structure.

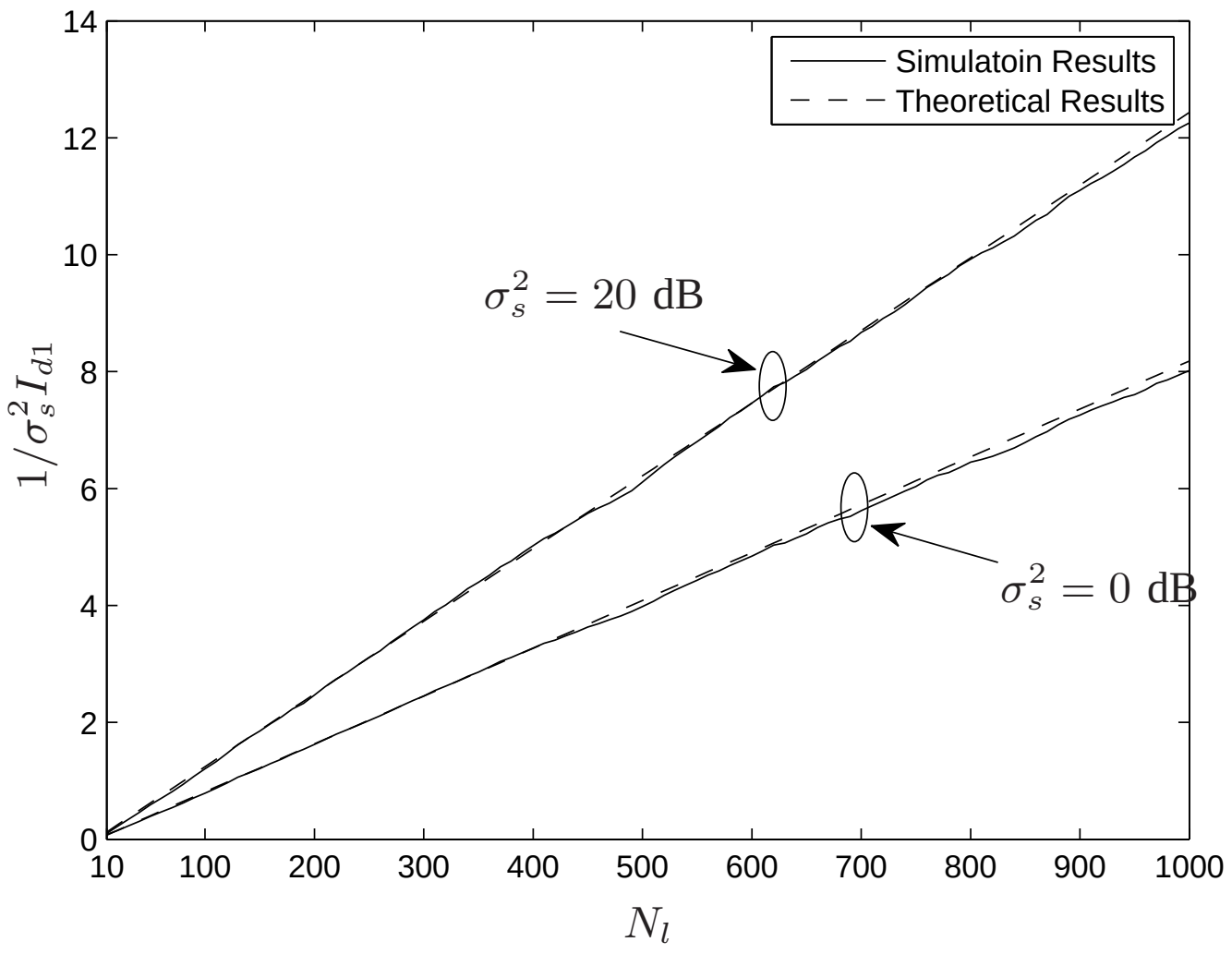

Fig. 3. Inverse of normalized IT versus environment learning time.

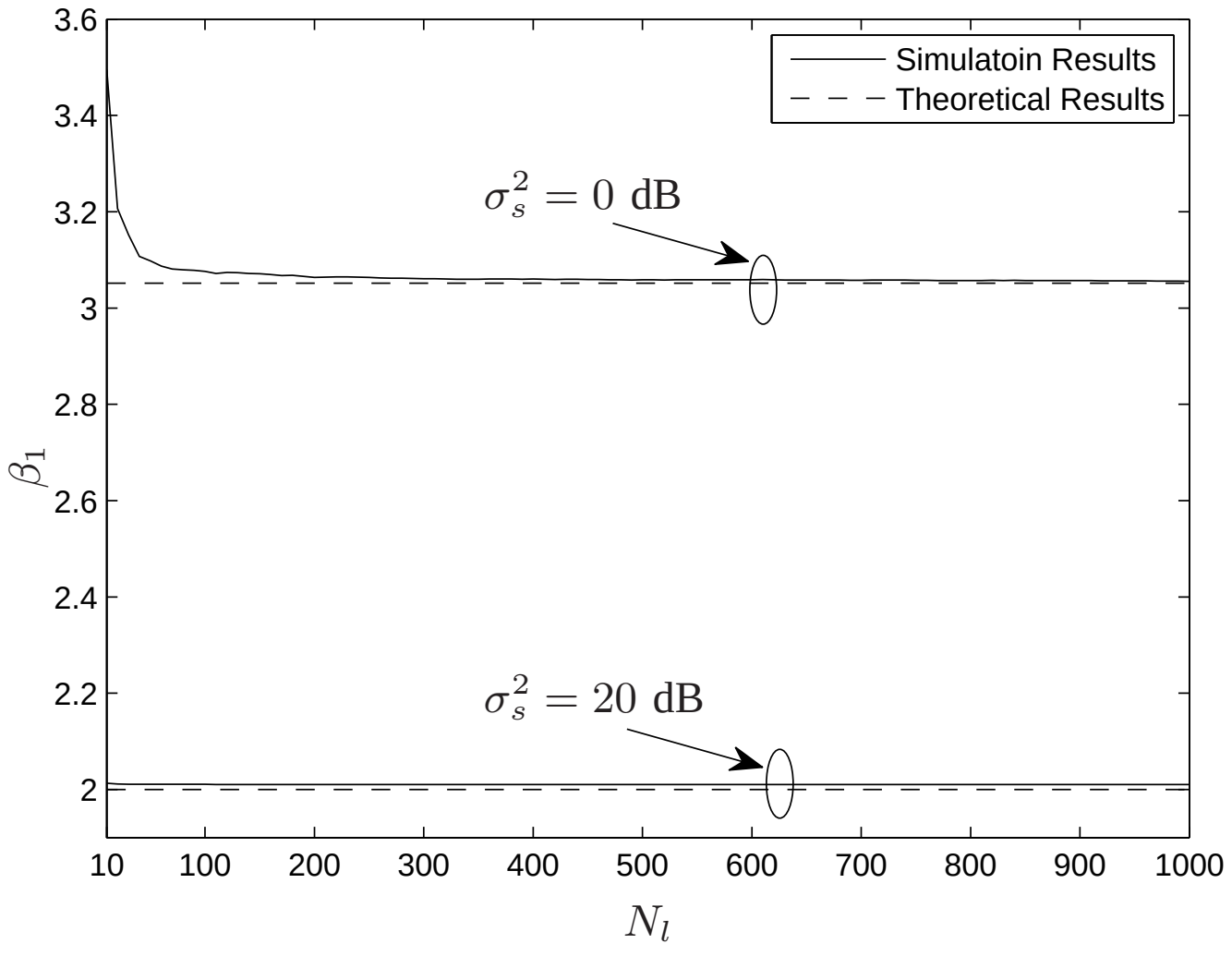

Fig. 4. The value of $\beta_1$ versus environment learning time.

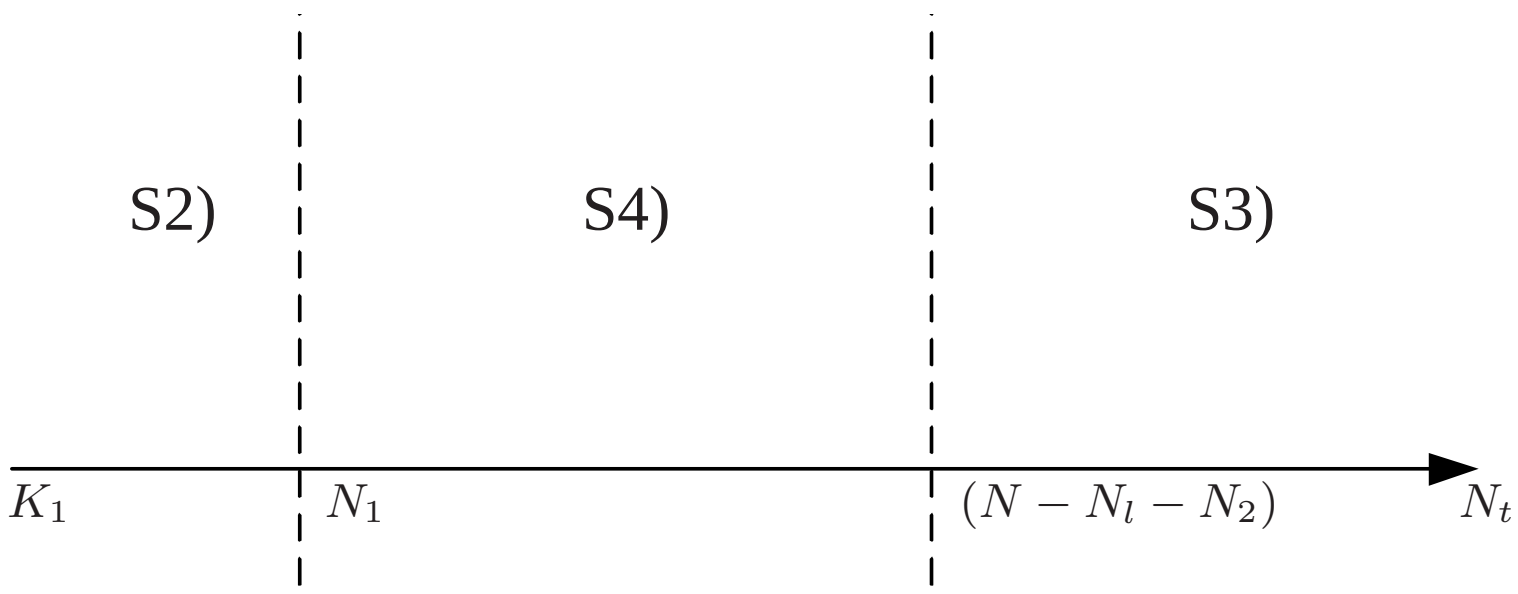

Fig. 5. The illustration of regions S2), S3), and S4), when $N_l \notin \mathcal{T}_l$.

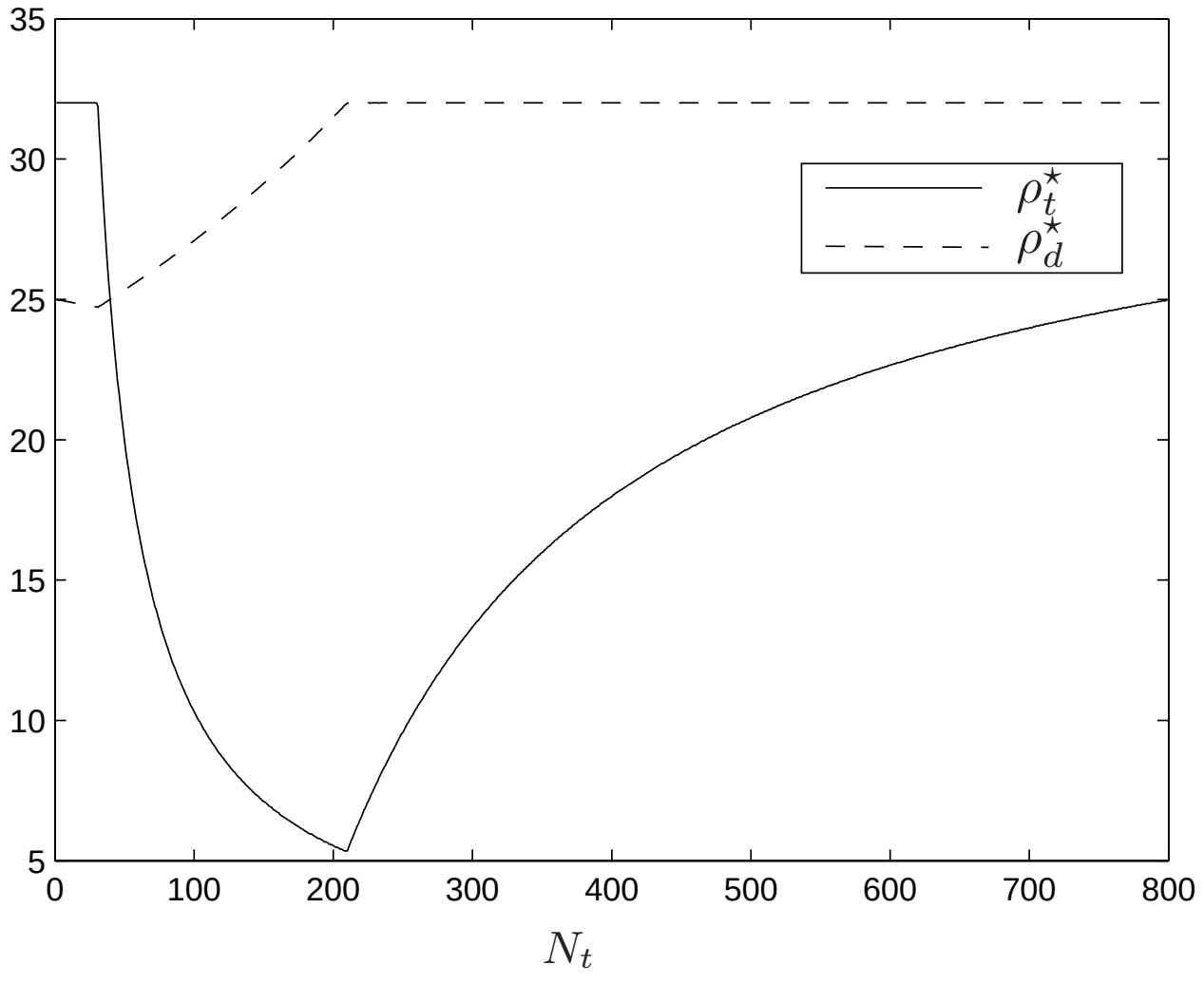

Fig. 6. The optimal powers $\rho_t^*$ and $\rho_d^*$ versus $N_t$ for $N_l \notin \mathcal{T}_l$.

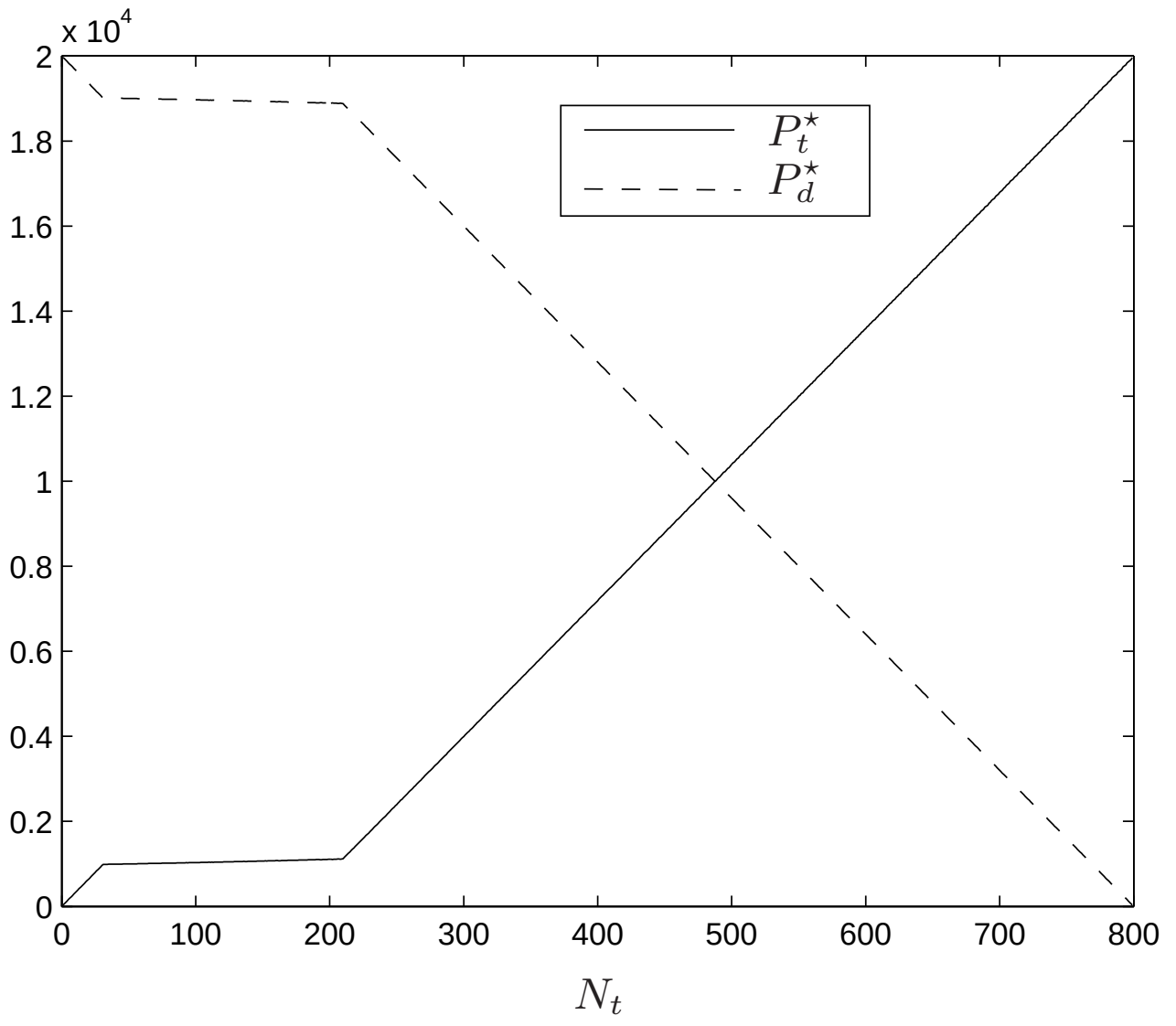

Fig. 7. The optimal total training power $P_t^\star$ and data transmission power $P_d^\star$ versus $N_t$ for $N_l \notin \mathcal{T}_l$.

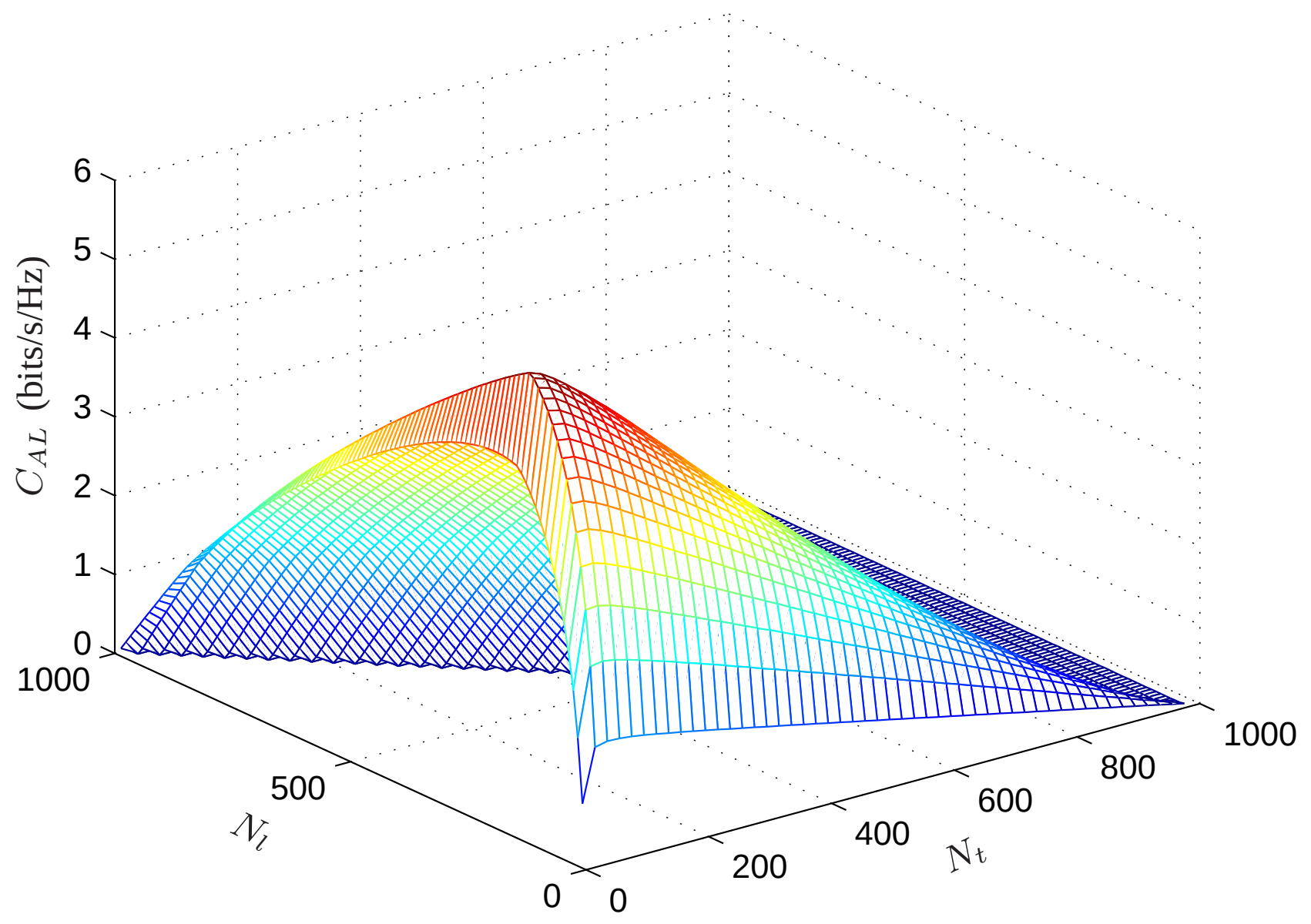

Fig. 8. $C_{AL}$ versus $N_l$ and $N_t$ with optimal power allocation, $\chi_1 = 0.16$.

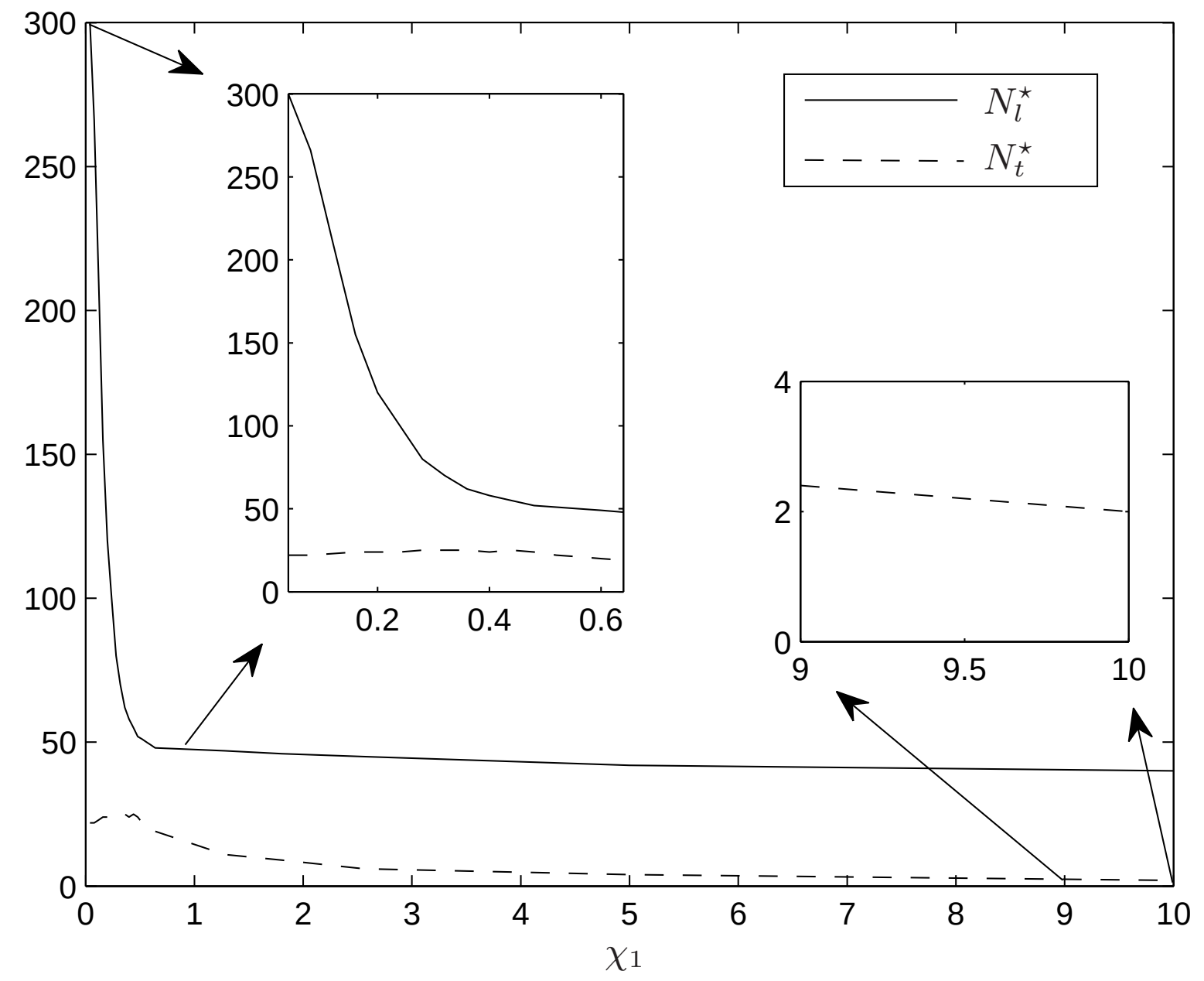

Fig. 9. Optimal $N_l$ and $N_t$ versus $\chi_1$.

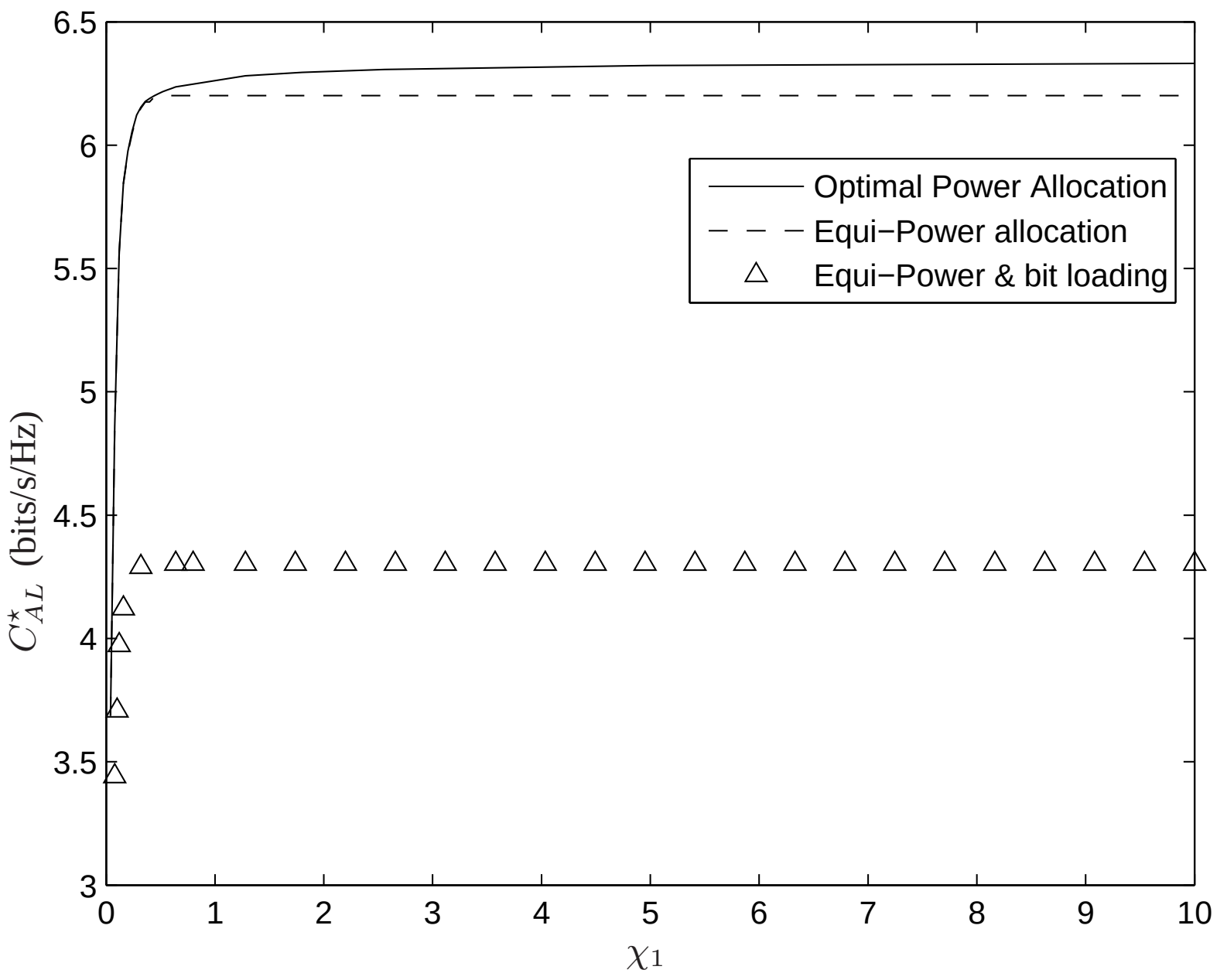

Fig. 10. $C_{AL}^\star$ versus $\chi_1$.